\documentclass[fleqn,usenatbib]{mnras}

\usepackage{newtxtext,newtxmath}
\usepackage{longtable}
\usepackage{multirow}

\usepackage[T1]{fontenc}

\DeclareRobustCommand{\VAN}[3]{#2}
\let\VANthebibliography\thebibliography
\def\thebibliography{\DeclareRobustCommand{\VAN}[3]{##3}\VANthebibliography}

\usepackage{graphicx}	
\usepackage{amsmath}	

\newcommand{\emerge}{{\sc~emerge} }

\newcommand{\Mpc}{\ensuremath{\rm~Mpc}}
\newcommand{\kpc}{\ensuremath{\rm~kpc}}

\newcommand{\Msun}{\ensuremath{{\rm M}_{\odot}}} 
 
\newcommand{\kms}{\ensuremath{{\rm~km\,s}^{-1}}}

\newcommand{\lmstar}{\ensuremath{\log_{10}(m^*/\Msun)}}
\newcommand{\lcdm}{\ensuremath{\Lambda}CDM }
\newcommand{\lpmvir}{\ensuremath{\log_{10}(M^{\mathrm{peak}}_{\mathrm{h}}/\Msun)}}

\newcommand{\dex}{\ensuremath{{\rm~dex}}}

\title[Dwarf SHMR]{Predictions on the stellar-to-halo mass relation in the dwarf regime using the empirical model for galaxy formation \sc Emerge}

\author[J. A. O'Leary et al.]{
Joseph A. O'Leary,$^{1}$\thanks{E-mail: joleary@usm.lmu.de}
Ulrich P. Steinwandel,$^{2}$\thanks{E-mail: usteinwandel@flatironinstitute.org}
Benjamin P. Moster,$^{3}$
Nicolas Martin,$^{4}$
Thorsten Naab$^{3}$
\\
$^{1}$Universit{\"a}ts-Sternwarte, Ludwig-Maximilians-Universit{\"a}t M{\"u}nchen, Scheinerstr. 1, 81679 M{\"u}nchen, Germany\\
$^{2}$Center for Computational Astrophysics, Flatiron Institute, 162 5th Avenue, New York, NY 10010\\
$^{3}$Max-Planck Institut f{\"u}r Astrophysik, Karl-Schwarzschild Stra{\ss}e 1, 85748 Garching, Germany\\
$^{4}$Observatoire Astronomique de Strasbourg, Universit{\'e} de Strasbourg, 11 rue de l'Universit{\'e}, F-67000 Strasbourg, France
}

\date{Accepted XXX. Received YYY; in original form ZZZ}

\pubyear{2022}

\begin{document}
\label{firstpage}
\pagerange{\pageref{firstpage}--\pageref{lastpage}}
\maketitle

\begin{abstract}
One of the primary goals when studying galaxy formation is to understand how the luminous component of the Universe, galaxies, relates to the growth of structure which is dominated by the gravitational collapse of dark matter haloes. The stellar-to-halo mass relation probes how galaxies occupy dark matter haloes and what that entails for their star formation history. We deliver the first self-consistent empirical model that can place constraints on the stellar-to-halo mass relation down to log stellar mass $\lmstar \leq 5.0$ by fitting our model directly to Local Group dwarf data. This is accomplished by penalising galaxy growth in late-forming, low-mass haloes by mimicking the effects of reionization. This process serves to regulate the number density of galaxies by altering the scatter in halo peak mass $M^{\mathrm{peak}}_{h}$ at fixed stellar mass, creating a tighter scatter than would otherwise exist without a high-$z$ quenching mechanism. Our results indicate that the previously established double-power law stellar-to-halo mass relation can be extended to include galaxies with $\lpmvir\gtrsim 10.0$. Furthermore, we show that haloes with $\lpmvir\lesssim 9.3$ by $z=4$ are unlikely to host a galaxy with $\lmstar > 5.0$. 
\end{abstract}

\begin{keywords}
galaxies: abundances, dwarf, Local Group, mass function, haloes, formation
\end{keywords}



\section{Introduction}
\label{sec:theory}
Galaxy formation in the $\Lambda$CDM framework predicts that dwarf galaxies are expected to be the most abundant galaxies in the universe, however their low luminosities make them particularly difficult to observe in practice. Meanwhile their sensitivity to feedback processes makes them difficult to model. Additionally, their shallow gravitational potential wells make them not only sensitive to internal feedback processes but also to assumptions on cosmology. Together this makes dwarf galaxies one of the best test-beds for our understanding of both cosmology and the fundamentals of galaxy formation.

Recent advancements in observational techniques have improved both the quantity and quality of dwarf galaxy observations. In particular these observations have probed to lower magnitudes offering data completeness to lower masses than previously available. Further followup measurements have advanced accuracy in measuring the distance \citep{Putman2021}, mass \citep{Woo2008} and star formation histories \citep{Weisz2014, Weisz2019} of these systems. These advancements present the possibility to better compare observations with high resolution theoretical models and open their use as direct constraining data for numerical models.

Most recent theoretical models have focused on utilising high resolution hydrodynamical zoom-in simulations to explore dwarf galaxies around the Milky Way \citep[e.g.][]{Sawala2015,Sawala2016b, GarrisonKimmel2019, Fattahi2020, Applebaum2021, Munshi2021}. These approaches simulate large cosmological volumes then re-simulate the Milky Way like haloes at higher resolution. The advantage of zoom simulations in studies of dwarfs is that it provides a large scale cosmological context while providing the resolution necessary to probe low mass satellites in systems similar to the Milky Way, where we have the most observational data. The draw back to this approach is that the SHMR can only be addressed within the context of Milky Way like haloes and generally do not explore how assumptions made impact the global SHMR. Furthermore, despite the improved resolution in hydrodynamical zooms, these models are still restricted by uncertain subgrid implementations that could impact the resulting dwarf population. 

Despite the fact that direct hydrodynamical zoom simulations are widely adopted in the field as a tool of choice to study dwarf galaxies around the Milky Way, there are some recent notable examples of work that also focused and introduced semi-numerical modelling to study dwarf galaxies \citep[e.g.][]{Kravtsov2022}.

Empirical models offer several distinct advantages over hydrodynamical and semi-analytic models. Because these models operate by relating galaxies to the host halo in post processing, computational power can instead be devoted to increasing mass resolution without the need to sacrifice statistics by simulating small volumes. And unlike either hydrodynamical simulations or semi-analytic models, empirical models make fewer assumptions on the relevant subgrid processes that impact galaxy formation. In this way the relationship between galaxies and haloes can be explored with less pollution from the personal priors imposed by subgrid models. The big caveat to empirical models is their need for observational data to constrain the model.

The lack of observational data has made empirical approaches to this problem difficult. There have been several recent attempts to quantify the stellar-to-halo mass relation (SHMR) using empirical techniques. \citet{Nadler2019} employed abundance matching on zoom-in simulations tuned to hydrodynamical simulations to make predictions on the abundance of low mass satellites down to $\lmstar \approx 2$. Meanwhile, using genetic algorithms \citet{Rey2022} explore how the merging histories of Milky Way mass hosts affects local dwarf population and statistics. Other recent works \citep{Wang2021} have taken an exploratory approach by extrapolating the \textsc{UniverseMachine} model \citep{Behroozi2019} into the ultra-faint regime. This is a useful technique to determine where the model must be improved to reproduce the observed characteristics of observed dwarfs. We expand on these approaches by using existing observations to directly constrain our own empirical model \emerge \citep{Moster2018}.

The goal of this work is to utilise real observables to constrain an empirical model that self consistently relates galaxy properties to dark matter halo properties at dwarf scales. Our aim is to better understand how low mass galaxies populate their haloes, and by doing so gain a better understanding of their star formation histories. This paper is organised as follows. First, in Sec.~\ref{sec:numerics} we will briefly introduce the $N$-body simulations that form the foundation of this work. We discuss the essential functionality and recent updates to the \emerge model in Sec.~\ref{sec:emerge_updates}. Additionally, in Sec.~\ref{sec:observation} we discuss the new observational data used to constrain our model. In Sec.~\ref{sec:rmodels} we introduce the model variations we explore in order to reproduce the observed properties of dwarf galaxies in the local Universe. Sec.~\ref{sec:results} compares our model implementations with one another and we discuss the resulting stellar to halo mass relation and star formation histories from our preferred model. Finally, in Sec.~\ref{sec:discussion} we discuss how our model assumptions might impact our results, in this context we discuss opportunities for future work and model improvements.

\section{Methods, Observations, Simulations etc.}
\label{sec:dwarf_sims}

\subsection{N-body simulations}
\label{sec:numerics}
\begin{figure*}
    \centering
    \includegraphics[width=0.97\columnwidth]{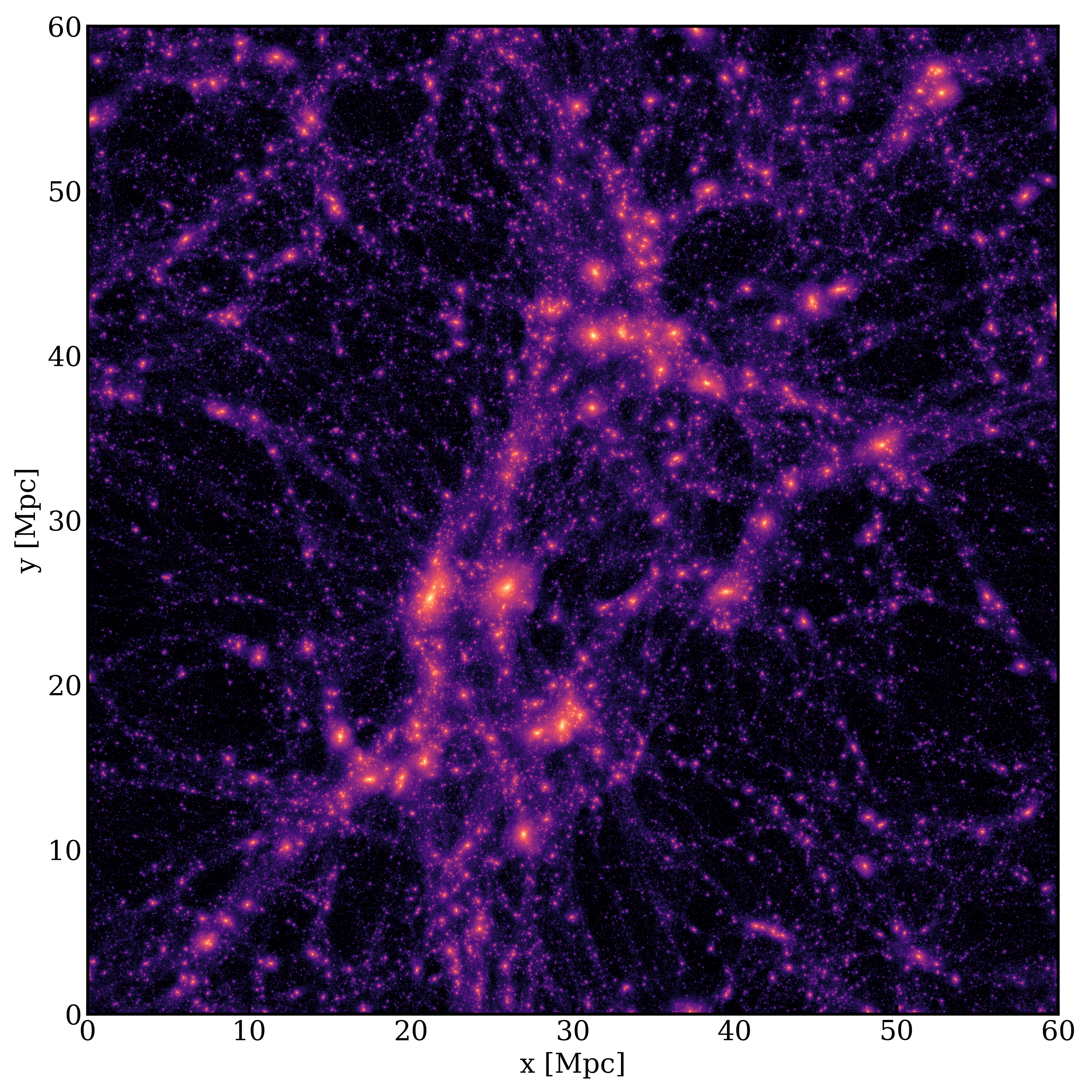}
    \includegraphics[width=\columnwidth]{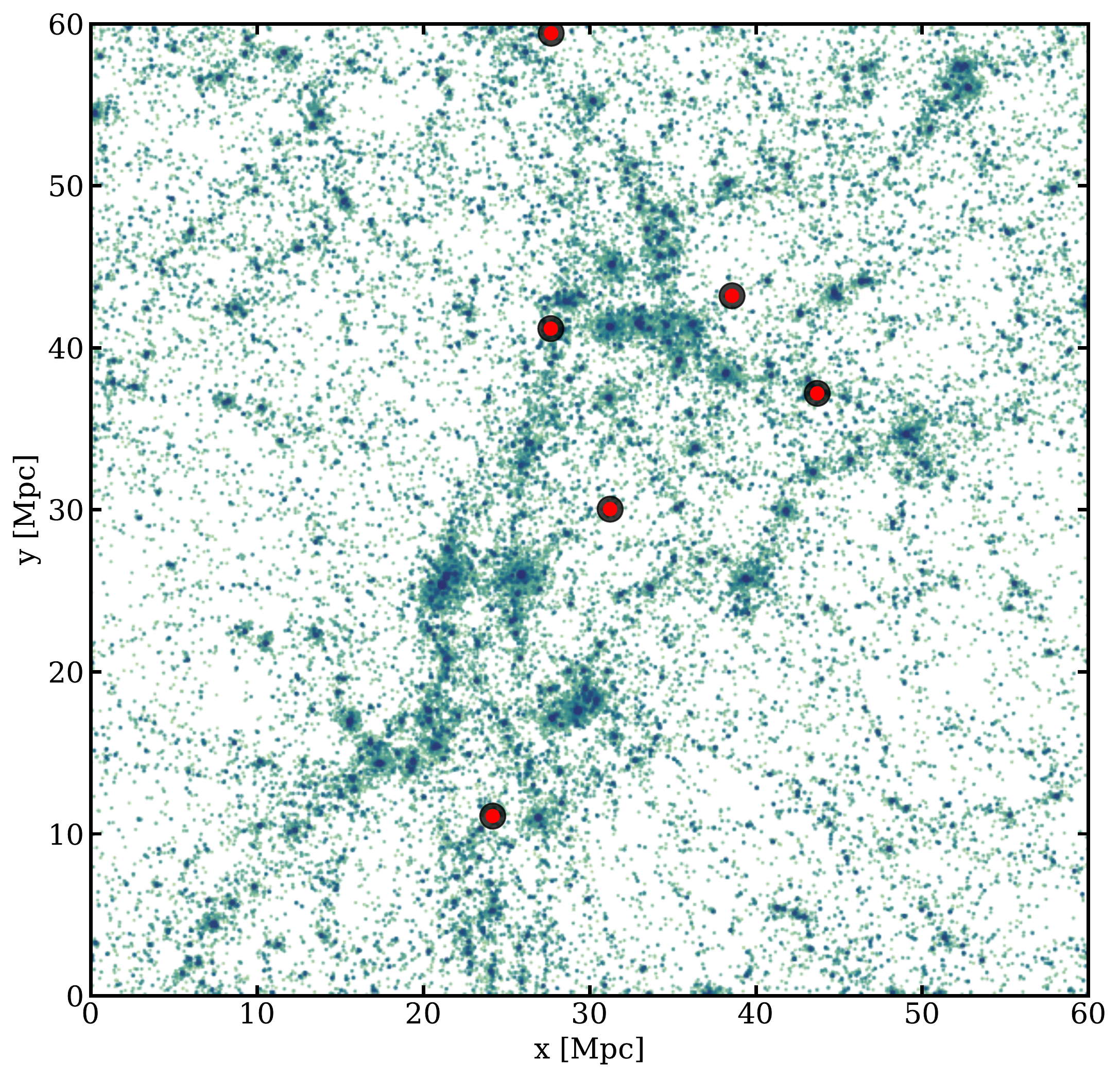}
    \caption{On the left, we show the dark matter surface density of our high resolution dark matter simulation with $2048^3$ particles, specifically designed to capture the smallest dark matter haloes accurately. On the right, we show \textit{all} central galaxies obtained with the empirical model \textsc{emerge}. Hereby, we highlight some of the Milky Way analogues in the log stellar mass range between 10.6 and 10.78 M$_{\odot}$ and the log halo mass range between 12 and 12.2 M$_{\odot}$, as black circles with a red face colour.} 
    \label{fig:royale}
\end{figure*}

We utilise a cosmological dark matter only $N$-body simulation in a periodic box with side lengths of $60$ Mpc. This simulation adopts $\Lambda$CDM cosmology consistent with \citep{PlanckCollaboration2016} results where $\Omega_m = 0.3080$, $\Omega_{\Lambda} = 0.6920$, $\Omega_b = 0.0484$, where $H_0 = 67.81\,\mathrm{km}\,\mathrm{s}^{-1}\mathrm{Mpc}^{-1}$, $n_s=0.9677$, and $\sigma_{8}=0.8149$. The initial conditions for this simulation were generated using \textsc{Music} \citep{Hahn2011} with a power spectrum obtained from \textsc{CAMB} \citep{Lewis2000}. The simulation contains $2048^3$ dark matter particles with particle mass $9.88\times10^5\mathrm{M}_{\odot}$. The simulation was run from $z=124$ to $0$ using the Tree-PM code \textsc{P-Gadget3} \citep{Springel2005b, Beck2016}. In total $147$ snapshots were created. Dark matter haloes are identified in each simulation snapshot using the phase space halo finder, \textsc{Rockstar} \citep{Behroozi2013}. Halo merger trees are constructed using \textsc{ConsistentTrees} \citep{Behroozi2013d}, providing detailed evolution of physical halo properties across time steps.\footnote{Additional information on these simulations, including: configuration files, build info, and parameter files can be found at \url{https://github.com/jaoleary}.} We show a density projection of the dark matter density in Fig.~\ref{fig:royale}.

\subsection{\emerge: just the basics}
\label{sec:emerge_updates}
In this section we provide a short overview of our empirical model \emerge. This model operates by populating galaxies in simulated dark matter merger trees by directly linking galaxy star formation to the individual growth histories of the host halo. The core of this approach is the baryon conversion efficiency which defines how effectively gas can be converted into stars at some halo mass: 
\begin{equation}\label{eq:eff}
    \epsilon(M) = 2\epsilon_{\mathrm{N}} \left[\left(\frac{M}{M_1}\right)^{-\beta} + \left(\frac{M}{M_1}\right)^{\gamma}\right]^{-1}
\end{equation}
where $\epsilon_{\mathrm{N}}$ is the normalisation, $M_1$ is the characteristic halo mass where peak efficiency occurs, $\beta$ specifies the efficiency slope at masses lower than $M_1$ and $\gamma$ is the slope at halo masses larger than $M_1$. The parameters are additionally allowed to scale linearly with scale factor such that:
\begin{eqnarray} \label{eq:ez}
	\log_{10} M_{1}(z) &=& M_0 + M_z\frac{z}{z+1} \;,\\
	\epsilon_N &=&  \epsilon_0 + \epsilon_z\frac{z}{z+1}\;, \\
	\beta(z) &=& \beta_0 + \beta_z\frac{z}{z+1}\;, \\
	\gamma(z) &=& \gamma_0 \;.
\end{eqnarray}
here parameters with subscript $0$ indicate the $z=0$ parameter values and subscript $z$ indicates the slope of scale factor evolution. The decision to allow parameters to evolve linearly is rather \textit{ad hoc}, that is there is no particular reason to chose a linear scaling other than its ability to reproduce observed statistics. 

For the work described in this paper we rely on recent improvements to the baryon conversion efficiency model in \emerge. Other recent works exploring the galaxy-halo connection with neural networks \citep{Moster2021} have shown that a linear-max scaling for $\epsilon_{\mathrm{N}}$ can provide an improved fit to observed stellar mass functions, especially at intermediate redshift. We have incorporated this proposed change into the version of \emerge used in this work. 
\begin{equation}
\epsilon_N=
\begin{cases} 
    \epsilon_0 + \epsilon_z\frac{z}{z+1} & \mathrm{if}\; \epsilon_N\leq \epsilon_{\mathrm{peak}} \\
    \epsilon_{\mathrm{peak}} & \mathrm{otherwise}
\end{cases}
\end{equation}
Here $\epsilon_{\mathrm{peak}}$ is a maximum allowed value for $\epsilon_N$ and is treated as an additional free parameter in the model.

\subsection{Observational data}
\label{sec:observation}

The empirical model is directly constrained by observational data. In addition to the data described in \citet{Moster2018} and \citet{OLeary2021} we extend the galaxy stellar mass function (SMF) data down to $\lmstar = 5$ through the inclusion of Local Group dwarf galaxies. In this work all galaxies within $2\Mpc$ are defined as members of the Local Group \citep{Putman2021}. We assign a galaxy as a satellite if it is positioned within $300\kpc$ of either the Milky Way or Andromeda (M31). We construct the SMF using the positively identified dwarf galaxies listed in online database of \citet{McConnachie2012}. Where available we use galaxy stellar masses from \citet{Woo2008} and a $1.6$ mass-to-light ratio otherwise \citep{Bell2001, Martin2008}. We assume $0.8\dex$ uncertainty in \lmstar for each system \citep{Woo2008}.

The dwarf mass function is constructed using $7$ bins evenly spaced in \lmstar. We then create $10^4$ random realisations of the locally observed dwarf population by sampling within the mass uncertainty range, each instance is then sampled again assuming the Poisson error of the distribution. From these random realisations we compute the average and $1\sigma$ interval mass function. Finally, we renormalise the dwarf mass function such that the average value matches the observed average SMF at $\lmstar=7$. This requires a vertical adjustment of $-0.83$ dex to the locally observed SMF. This vertical adjustment is necessary as the global mass function is computed over large volumes and incorporates more void space which lowers the average density compared to our locally measured mass function which focuses on our own over-dense region of the Universe.

Before moving on we should take a moment to consider the implications of relying on Local Group observations to characterise the dwarf galaxies in large volumes. The first clear draw back to this approach is due to the limited sample size. When probing down to $\lmstar=5$ observations are at best only complete out to $\sim2\Mpc$, this practical limitation restricts our sample of galaxies to $64$ confirmed objects. Although extended catalogues exist for dwarf satellites around other hosts and for even lower masses in the Local Group \citep{Geha2017, Smercina2018, Simon2018, Carlsten2020, Mao2021} these studies do not offer the the completeness we require to constrain our model. This low number of systems results in large error bars on the resulting SMF, which may reduce our ability to make a statistical distinction on the quality of the fit in our proposed model implementations. Nonetheless, we can still make arguments for, or against each model variation based on the resulting systems and whether their properties make sense with our current understanding of galaxy growth at low masses. Additionally, an argument could be made that locally observed trends are not representative of the universe at large; however this probes the current limit of our observational capabilities at small scales (see Section~\ref{sec:discussion}). Future observations from more sensitive instruments, such as JWST, will eventually allow us to explore more systems and build a more complete data set.

\section{High redshift quenching}
\label{sec:reionization}
In the discussion of dwarf galaxies there are two particular observational facts we need to contend with. The first is the abundance of dwarf galaxies. The $\Lambda$CDM paradigm predicts low mass haloes should appear in the greatest abundances, and recent studies of high redshift star formation indicate that haloes with masses as low as $M_{\mathrm{h}}\gtrsim10^5\Msun$ might be sufficient to initiate star formation \citep{Hirano2015, Schauer2021, Kulkarni2021}. If every low mass halo hosted a bright galaxy we would expect to observe many thousands of dwarf galaxies in the Local Volume. This prediction conflicts with the relatively low number of dwarf galaxies that have actually been observed and catalogued \citep{McConnachie2012}. The discrepancy between number of observed dwarfs, and the number predicted by $\Lambda$CDM is what is known as the missing satellite problem \citep[e.g.][]{Klypin1999,Moore1999}. 

The second complication is in the star formation history of these galaxies. Investigation of dwarfs in the Local Group reveals that these galaxies possess low star formation rates and tend to be much older. At the lowest masses observations indicate that for most stars formed by $z=6$ \citep{Weisz2014,Weisz2015}. This leaves us with the question: why are dwarf galaxies so old? 

One possible solution that can address both is that star formation efficiency is suppressed in low mass haloes at late times due to cosmic reionization. Early work investigating the impact of reionization on the abundance of dwarfs showed that including this feedback process in models for galaxy formation could reconcile the discrepancy of dwarf abundances between $\Lambda$CDM predictions and observation \citep[e.g.][]{Efstathiou1992, Thoul1996, Bullock2000, Somerville2002}. These models generally predicted that reionization inhibits star formation in haloes with a maximum circular velocity $V_{\mathrm{peak}}\lesssim20\kms$ ($M_{\mathrm{h}} \lesssim 10^9\Msun$).

In this work we build on these techniques by incorporating a model for high-$z$ quenching into \emerge. This model option will complement the already existing mechanisms that impact galaxies in low mass haloes such as environmental quenching and tidal disruption \citep[see][for details]{Moster2018, OLeary2021}.

\subsection{Model variations}
\label{sec:rmodels}
We tested three physically motivated models for a high-$z$ quenching process that impacts low mass haloes. Throughout this work we compare our results to the ``reference'' model which is our unaltered \textsc{emerge} model with the parameters shown in table~\ref{tab:mod_fit}. The reference case we show uses the same $60\Mpc$ box as described in Sec.~\ref{sec:dwarf_sims}. However, the parameters used were fit using a lower resolution $200\Mpc$ box \citep[described in][]{Moster2018}. The Local Group dwarf data was not incorporated into those fits. We also refit the standard \emerge model with the inclusion of Local Group data as an additional check, this variant is simply labeled \textit{Refit}. Additionally, all models were also fit without incorporating the lowest data point in the Local Group SMF, these model results are denoted by a suffice \textit{B} to the model name. This was done to address possible sample incompleteness at the lowest masses.

Our high-$z$ quenching models suppress star formation in low mass-high redshift haloes in a manner consistent with expectations from reionization. The shared characteristic of these methods is that they specify a minimum halo mass $M_{\mathrm{h}}^{\mathrm{min}}$ required to form stars at some scale factor $a=1/(1+z)$. When a halo does not meet that threshold its star formation will be set to zero and will remain zero for the remainder of that galaxy's lifetime. If a halo first appears in the simulation (leaf halo) below the specified threshold, no galaxy will be seeded and the halo will remain dark (see Sec.~\ref{sec:dwarf_futurework}).

The first and most simple model we test treats the quenching as a uniform, instantaneous process as described by eq.~\ref{eq:inst}. Here star formation is shut off in haloes of insufficient mass $M_{\mathrm{q}}$ by the specified scale factor $a_{\mathrm{q}}$. We linearly interpolate halo masses between snapshots to avoid imprinting a preferred quenching scale factor due to time-step discreteness.
\begin{equation}
\label{eq:inst}
M_{\mathrm{h}}^{\mathrm{min}}(a)=
\begin{cases} 
        M_{\mathrm{q}} & \mathrm{if}\; a > a_{\mathrm{q}}, \\
        0 & \mathrm{otherwise}.
   \end{cases}
\end{equation}
Both of these parameters are free in the model. This model is referred to as \textit{instantaneous} for the remainder of the paper.

Our next model is a linear-max construction given by eq.~\ref{eq:linmax}. This describes a process where the minimum halo mass needed to form stars increases linearly with increasing scale factor up to some maximum scale factor, after which the halo mass threshold remains constant.
\begin{equation}
\label{eq:linmax}
M_{\mathrm{h}}^{\mathrm{min}}(a)=
\begin{cases} 
        R_{\mathrm{q}}(a-a_{\mathrm{q}}) + M_{\mathrm{q}} & \mathrm{if}\; a\leq a_{\mathrm{q}}, \\
        M_{\mathrm{q}} & \mathrm{otherwise}.
   \end{cases}
\end{equation}
Here $R_{\mathrm{q}}$ indicates the rate at which the mass threshold increases with scale factor and $a_{\mathrm{q}}$ indicates the scale factor where the threshold reaches its maximum $M_{\mathrm{q}}$. This model is refereed to as \textit{lin-max} hereafter.

Finally, we test a logistic model which allows for continuously increasing reionization threshold up to some maximum defined by eq.~\ref{eq:logistic}.
\begin{equation}
\label{eq:logistic}
    M_{\mathrm{h}}^{\mathrm{min}}(a) = \frac{M_{\mathrm{q}}}{1+\exp\left[-R_{\mathrm{q}}(a-a_{\mathrm{q}})\right]}\, ,
\end{equation}
here $M_{\mathrm{q}}$ is the maximum threshold mass, $R_{\mathrm{q}}$ is the transition strength, and $a_{\mathrm{q}}$ is the midpoint scale factor where the rate reaches its maximum. We implement eq.~\ref{eq:logistic} in log-space such that the minimum mass threshold is $10^0 \Msun$. Our simulation is of insufficient resolution to probe galaxy formation on those scales so this floor is somewhat artificial. If the floor value is changed we could anticipate possible changes in $R_{\mathrm{q}}$, $a_{\mathrm{q}}$ or both. While this formulation contributes the same number of free parameters as lin-max, the smooth continuous transition delivers a more physically interpretable view of the reionization process, while also offering greater flexibility to incorporate additional parameters if needed. This model will be referred as \textit{logistic} for the remainder this work.
Furthermore, adopting an evolving threshold mass is consistent with \citet{Ricotti2008, Rey2020, Benitez-Llambay2021} who employ en evolving UV background during the epoch of reionization.

\subsection{Fitting}
When performing an $N$-body simulation there is typically a trade off between the size of the simulation volume and the particle resolution. Our simulation parameters were chosen to maximise the number of Milky Way like systems while simultaneously providing the resolution necessary to capture the stellar mass to halo mass relation \textit{and} the associated scatter down to $\lmstar=5$. This decision, while necessary for this study, limits which model parameters can be explored in the fitting process largely due to the limited number of more massive systems. Subsequently we only allow the low mass slope of the baryon conversion efficiency $\beta_0$, its redshift evolution $\beta_z$ and the stellar mass dependent quenching timescale $\tau_\mathrm{s}$ as free parameters in addition to those introduced by our new model variations. All remaining free parameters of the model are fixed to values indicated in table~\ref{tab:mod_fit}. Additionally, we utilised binned observational data as opposed to the raw observations as was done in \citet{Moster2018}. By fitting to binned data we minimise the chance of the MCMC getting stuck in a local minimum due to conflicting observations. Finally, we apply a prior to the binned data. This prior applies increased weight to data at low-$z$ and near strong inflection points in order to preserve observed trends where the data is most robustly measured\footnote{The data used for fitting along with associated weights can be obtained at \url{https://bitbucket.org/bmoster/emerge}.}.

\section{Results}
\label{sec:results}
\subsection{Fits and model selection}
Here we discuss how each model performs in reproducing the range of the observed data, as well as evaluate some evidence based model selection criteria. Tab.~\ref{tab:mod_fit} shows the best fit parameters for each model. Empty fields indicate that the parameter is either free or not relevant to the model, otherwise the parameters for the reference model have been used. The most notable change to the model parameters is the low mass baryon conversion efficiency slope $\beta_0$. With the extended SMF all models tend to prefer a steeper conversion slope indicating less effective star formation compared to the reference.
\begin{table}
	\centering
	\caption[Best fit model parameters for each high redshift quenching model variant]{The best fit model parameters for each model variant used in this work. empty fields indicate the parameter was not left free when fitting that model or was not available in that model. Noted confidence intervals correspond to the $\pm 1\sigma$ range.}
	\label{tab:mod_fit}
	\begin{tabular}{lcccr}
		\hline
		Parameter    & Reference  & Instantaneous & Lin-max & Logistic \\
		\hline
		\hline
		$M_0$        & $11.32^{+ 0.03}_{-0.02}$  & - & - & -  \\
		$M_z$        & $1.45^{+0.06}_{-0.07}$  & - & - & -  \\
		$\epsilon_0$ & $0.02^{+0.01}_{-0.01}$  & - & - & -  \\
		$\epsilon_z$ & $1.70^{+0.01}_{-0.02}$  & - & - & -  \\
		$\epsilon_{\mathrm{peak}}$ & $0.30^{+0.02}_{-0.01}$  & - & - & -  \\
		$\beta_0$    & $2.98^{+0.20}_{-0.22}$  & $2.37_{-0.02}^{+0.03}$ & $1.89_{-0.02}^{+0.04}$ & $2.22_{-0.05}^{+0.09}$  \\
		$\beta_z$    & $-2.68^{+0.26}_{-0.22}$ & $-1.70_{-0.05}^{+0.03}$ & $-1.18_{-0.05}^{+0.02}$ & $-1.50_{-0.11}^{+0.04}$  \\
		$\gamma_0$   & $1.25^{+0.01}_{-0.01}$  & - & - & -  \\
		\hline
		$f_{esc}$    & $0.59^{+0.02}_{-0.02}$  & - & - & -  \\
		$f_{s}$      & $0.01^{+0.01}_{-0.01}$  & - & - & -  \\
		$\tau_{0}$   & $0.81^{+0.10}_{-0.18}$  & - & - & -  \\
		$\tau_{s}$   & $0.52^{+0.06}_{-0.05}$  & $0.40_{-0.01}^{+0.01}$ & $0.42_{-0.01}^{+0.01}$ & $0.40_{-0.01}^{+0.02}$  \\
		\hline
		$a_{\mathrm{q}}$ & - & $0.23_{-0.02}^{+0.03}$ & $0.24_{-0.03}^{+0.03}$ & $0.19_{-0.03}^{+0.03}$\\
		$M_{\mathrm{q}}$ & - & $9.40_{-0.07}^{+0.09}$ & $9.34_{-0.19}^{+0.05}$ & $9.33_{-0.08}^{+0.11}$\\
		$R_{\mathrm{q}}$ & - & - & $1.14_{-0.06}^{+0.06}$ & $2.56_{-0.52}^{+0.38}$\\
		\hline
	\end{tabular}
\end{table}
\begin{figure}
	\includegraphics[width=\columnwidth]{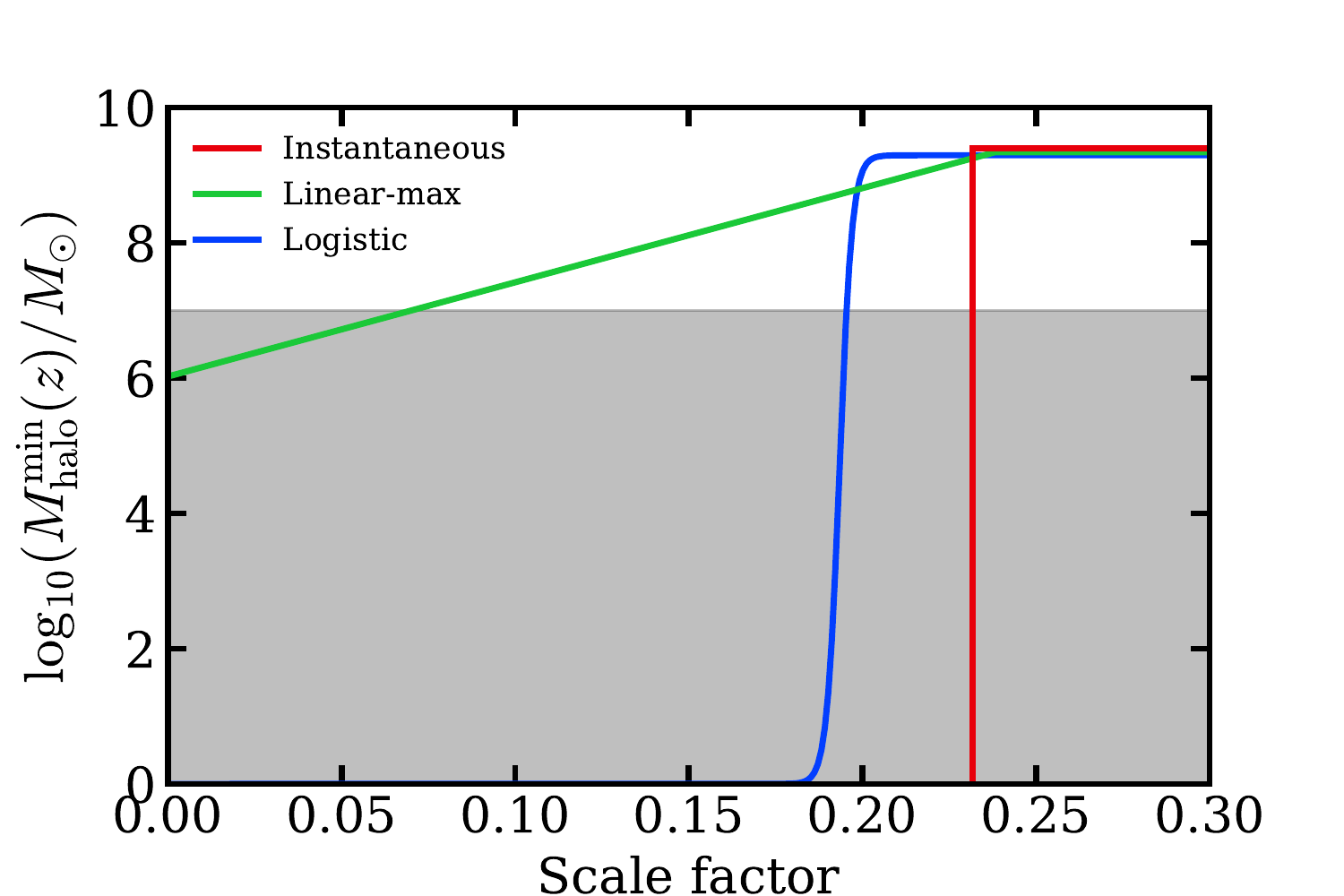}
    \caption[Quenching mass]{The evolution of $M_{\mathrm{h}}^{\mathrm{min}}(a)$ under our three model variations. Haloes with mass below $M_{\mathrm{h}}^{\mathrm{min}}(a)$ (coloured lines) have star formation instantaneously quenched. The grey region blocks our sub-resolution halo masses.}
	\label{fig:mod_params}
\end{figure}
All three of our model variations converge to $M_{\mathrm{q}}\approx9.3-9.4$. With the quenching mass saturating between $z\approx3.4-3.75$. The lin-max and logistic model options, which allow for a time dependent $M_{\mathrm{h}}^{\mathrm{min}}$, do not agree on the rate of increase. These models nonetheless show nearly identical reproduction of the observational data used to constrain \emerge as well as in the resulting SHMR and star formation histories of dwarfs (see Sec.~\ref{sec:dwarf_shmr} and Sec.~\ref{sec:dwarf_sfh} respectively). This likely means there is insufficient data to constrain the time evolution of $M_{\mathrm{h}}^{\mathrm{min}}$ and that it is not \textit{required} to explore galaxies with $\lmstar > 5$.

\begin{figure}
	\includegraphics[width=\columnwidth]{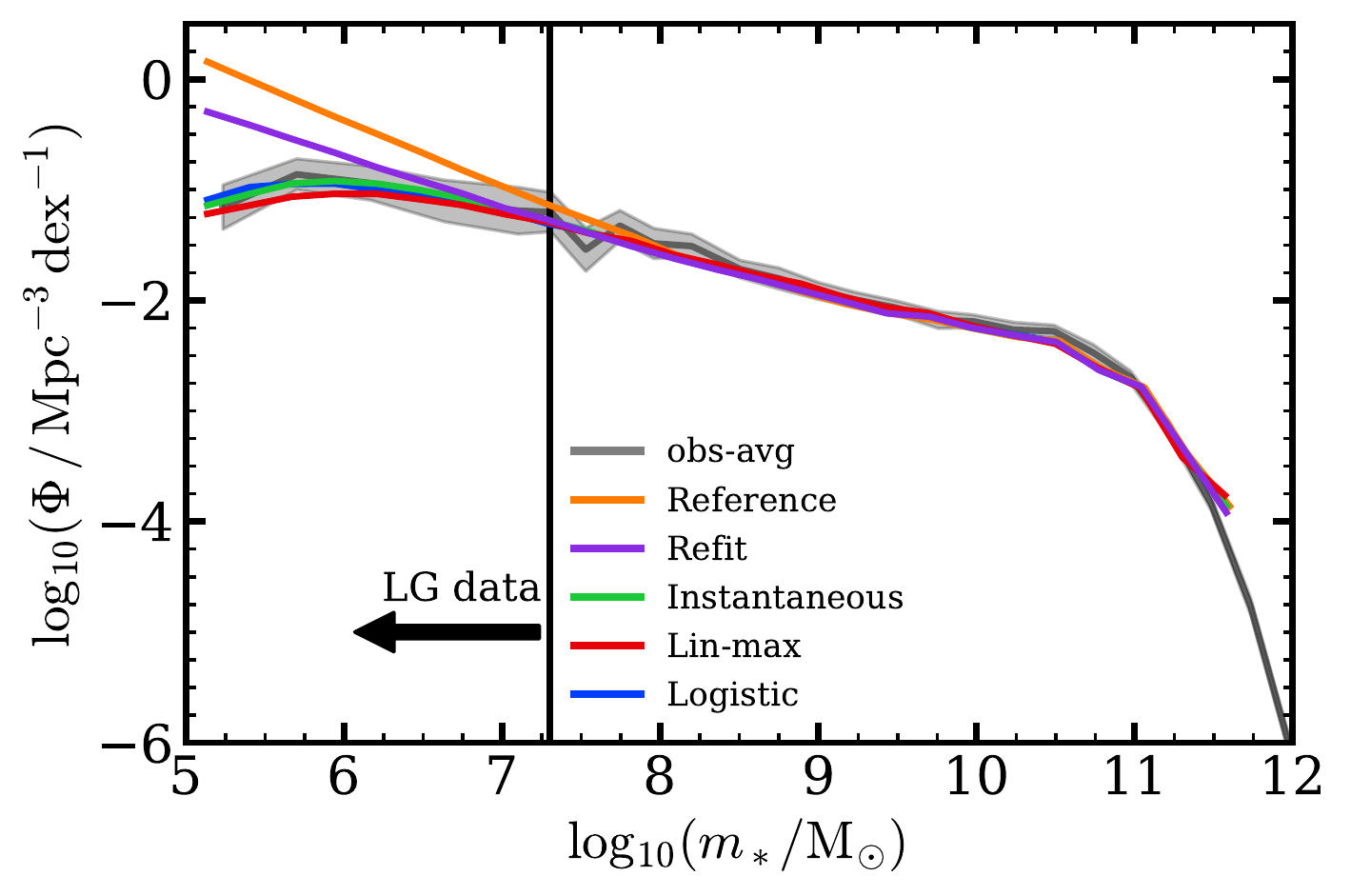}
    \caption[Impact of high redshift quenching on the global stellar mass function]{A comparison of the global galaxy stellar mass function for our model variations against the observed average stellar mass function. Solid lines illustrate the resulting $z=0$ stellar mass function using best fit parameters for each model variant. The grey line and shaded region indicate the observed SMF average from a suite of observational estimates along with the $68$ per cent confidence interval. The observed data to the left of the vertical black line illustrates the region where we have extended the observed SMF using the Local Group dwarfs listed in \ref{tab:obsdat}}
	\label{fig:GSMF}
\end{figure}

Fig.~\ref{fig:GSMF} shows how each of our models reproduces the $z=0$ mass function (coloured lines), compared with the reference (orange line). Each of our model variations successfully reproduces relevant observables, and qualitatively no distinction can be visually identified with respect to the SMF. The lower star formation efficiency in the new models does however result in a lower density in the $7\leq \lmstar < 8$ range, but all remain consistent with observed data.

A more quantitative comparison can be performed using an information criterion\footnote{Details on these methods are described in the appendix of \citet{Moster2018}}. This way we can weight the quality of the fit for each model against the added complexity of additional free parameters. Higher order models can fit the data better, but at some point there will not be enough information to further constrain additional parameters and increasing the order will not provide a better fit to the data. In these schemes models are penalised as additional parameters are added, providing a pathway to selecting the most simple model that can reproduce the data. Tab.~\ref{tab:mod_stats} lists the statistical characteristics of each model.
\begin{table*}
	\centering
	\caption[High redshift quenching model statistics]{Model statistics. Models denoted `B' only include data points $\log_{10}(m_*/M_{\Msun})>6.0$}
	\label{tab:mod_stats}
    \begin{tabular}{lcccccccr} 
		\hline
		Model    & $\chi^{2}_{\mathrm{min}}$  & $\chi^{2}_{\mathrm{mean}}$ & $N_{\rm p}$ & $p_{\rm D}$ & AIC & BIC & DIC & $-2\ln(Z)$\\
		\hline
		\hline
		Refit           & 906.07 & 910.69 & 3 & 4.62 & 912.15 & 923.34 & 915.31 & 933.63\\
		Refit\_B        & 790.56 & 792.88 & 3 & 2.32 & 796.64 & 807.82 & 795.20 & 815.28\\
        Instantaneous   & 822.38 & 826.59 & 5 & 4.20 & 832.58 & 851.16 & 830.79 & 858.36\\
        Lin-max         & 781.83 & 788.06 & 6 & 6.23 & 794.10 & 816.37 & 794.29 & 828.48\\
        Logistic        & 827.08 & 830.00 & 6 & 2.93 & 839.35 & 861.61 & 832.93 & 860.24\\
        Logistic\_B     & 823.02 & 826.23 & 6 & 3.21 & 835.29 & 857.53 & 829.44 & 846.45\\
		\hline
	\end{tabular}
\end{table*}

From a statistical point of view the model lin-max provides the best reproduction of observed data. Beyond these statistical measures we find very little quantitative differences between each model. For the remainder of this work we will make all of our comparisons with respect to the logistic model. While the logistic model does not provide any additional predictive power with respect to lin-max, the logistic model provides the greatest flexibility for incorporating new observables, along with the possibility to set a halo mass floor for galaxy formation. Rather than selecting a model which will need to be fundamentally overhauled given new data, it may be preferable to select the model which readily accepts additional parameters to increase its complexity when needed. Ultimately, we found that our conclusions are unaffected by the choice of the high-$z$ quenching model variation we adopt. Furthermore, although refitting the reference with the included local group data provides a better fit, we elect to make our comparisons to the unaltered \emerge parameters. Our comparisons then occupy two extremes to provide a more clear contrast on how the inclusion of this data and high-$z$ quenching impact the SHMR and growth history of dwarfs. Although we will highlight the results from only two models, figures illustrating how each model variant agrees with observed constraints, as well as predicted relationships and galaxy properties are shown in \ref{sec:appB}.

\subsection{The stellar-to-halo mass relation}\label{sec:dwarf_shmr}
The primary purpose of this work is not to explore the physics of reionization, but to better understand how observed dwarfs came to be, and how these low mass objects fit into the \lcdm paradigm. In particular we are interested in the relationship between dwarf galaxies and their dark matter haloes and the associated scatter in that relationship.

\begin{figure*}
	\includegraphics[width=\textwidth,height=\textheight,keepaspectratio]{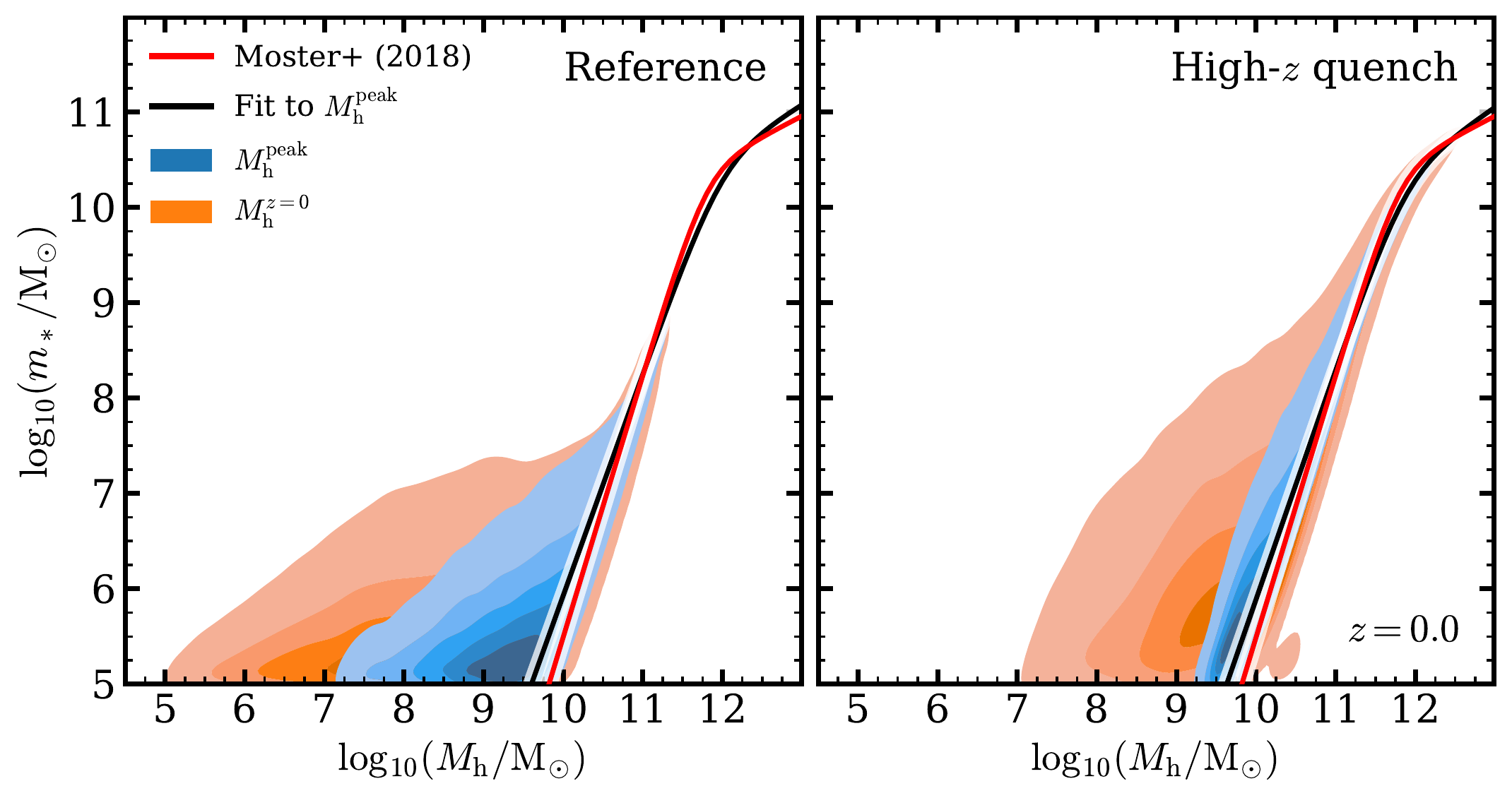}
    \caption[The stellar-to-halo mass relation in the dwarf regime]{The stellar-to-halo mass relation under the reference model (left panel) compared with the logistic high-$z$ quenching model (right panel).  The blue and orange contours show the iso-proportion contours of the SHMR in terms of halo peak mass $\log_{10}(M_{\mathrm{h}}^{\mathrm{peak}}/\mathrm{M}_{\odot})$ and present day halo mass $\log_{10}(M_{\mathrm{h}}^{z=0}/\mathrm{M}_{\odot})$ respectively. The solid lines indicate the best fit SHMR assuming a double-power law relation. The black lines shows the average relation from the data in this work while the red lines show the relation from \citet{Moster2018}.}
	\label{fig:shmr_join}
\end{figure*}
Here we discuss how the SHMR from our preferred high-$z$ quenching model compares with the reference case. Fig.~\ref{fig:shmr_join} shows as side by side comparison of these two models in terms of halo peak mass $\log_{10}(M_{\mathrm{h}}^{\mathrm{peak}}/\mathrm{M}_{\odot})$ (blue contours) and present day halo mass $\log_{10}(M_{\mathrm{h}}^{z=0}/\mathrm{M}_{\odot})$ (orange contours). \citet{Moster2010} shows that galaxy stellar mass at any epoch is closely related to the peak halo mass of that galaxy, and the resulting SHMR can be well fit by a double power-law. We can see from the black solid lines in Fig.~\ref{fig:shmr_join} that the average SHMR can be well approximated by a power law relation down to at least $\lpmvir \approx 10$. The results of these models is consistent with an extrapolation of the power-law relation (solid red lines) shown in \citet{Moster2018}.

How this relationship evolves for $\lpmvir \lesssim 10$ is difficult to ascertain. The reference model produces a relation that extends as a power-law to lower masses. In this case the relationship is primarily determined by the low mass slope of the baryon conversion efficiency. Here, altering the low mass conversion slope can change the slope but the SHMR remains a power-law, by construction. Introducing high-$z$ quenching breaks this relationship by preventing some haloes from forming galaxies in a way that is proportional to the halo growth. This can be clearly seen in the blue contours of the high-$z$ quenching model (right panel) of Fig.~\ref{fig:shmr_join}. Here we can see a rapid cut off in the SHMR as we approach  $M_{\mathrm{q}}$. Comparing the blue contours of the reference (left panel) and high-$z$ quenching models we can see that although the models share a similar average SHMR, they diverge strongly in the scatter of \lpmvir at fixed \lmstar. We can conclude that the suppression of the SMF at low masses is primarily due to the elimination of late forming low mass ratio systems, $m_*/M^{\mathrm{peak}}_{\mathrm{h}}$, due to a high redshift quenching process.

\begin{figure*}
	\includegraphics[width=\textwidth,height=\textheight,keepaspectratio]{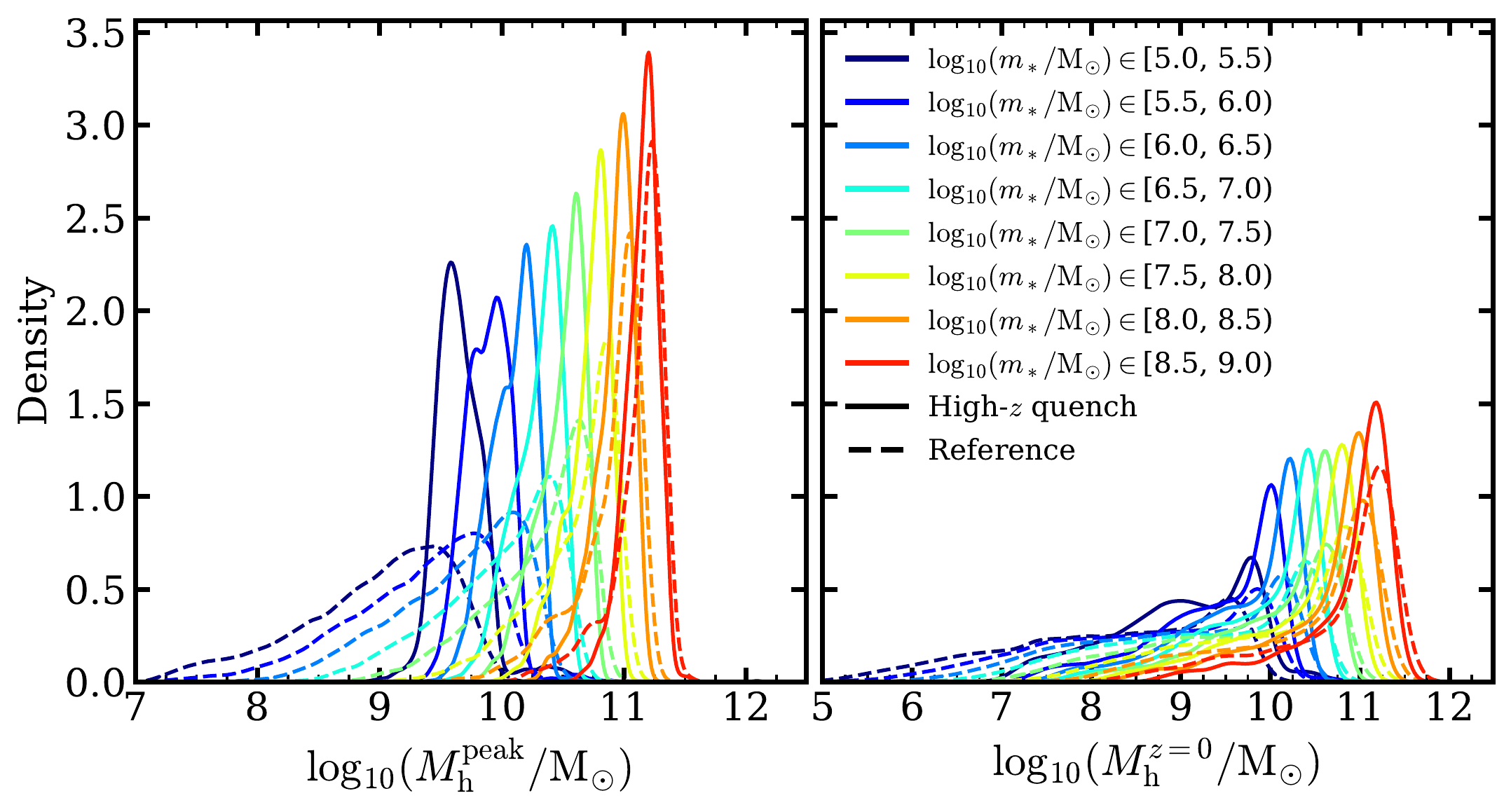}
    \caption[The distribution of halo masses at fixed stellar mass in the dwarf regime]{The distribution of halo masses at fixed stellar mass. Solid lines correspond to the logistic model and dashed lines are for the reference model. Colours indicate the mass bin for each distribution. The left panel shows distributions in peak halo mass $\log_{10}(M_{\mathrm{h}}^{\mathrm{peak}}/\mathrm{M}_{\odot})$ and the right panel shows distributions in present day halo mass $\log_{10}(M_{\mathrm{h}}^{z=0}/\mathrm{M}_{\odot})$.}
	\label{fig:shmr_scatter}
\end{figure*}
Previous work has shown that the scatter in the SHMR takes a log-normal distribution at a fixed halo mass \citep{Cooray2006}, similarly we find that our results exhibit this same trend down to $\lpmvir \approx 10$. Due to the lack of observable constraints for $\lmstar<5$ we are unable to verify that this trend extends to lower halo masses. Instead we can compare the scatter in \lpmvir at fixed stellar mass to better evaluate how high-$z$ quenching impacts scatter compared to the reference treatment.

Fig.~\ref{fig:shmr_scatter} illustrates the scatter in halo mass at fixed stellar mass for \lpmvir\; (left panel) and $\log_{10}(M_{\mathrm{h}}^{z=0}/\mathrm{M}_{\odot})$ (right panel). We can see that down to $\lmstar \approx 7$ the distribution in \lpmvir\; remains approximately Gaussian with a slight tail extension toward lower masses. For $\lmstar \lesssim 7$ we find that the standard implementation (dashed lines) produces a highly asymmetric distribution with a pronounced tail toward low halo masses. Conversely, introducing high-$z$ quenching (solid lines) not only reduces the range in scatter but results in a \textit{more} Gaussian distribution. Looking at the same distributions with respect to present day halo mass (right panel) we can see that in the reference model low-mass galaxies occupy a much larger range of halo mass than in the high-$z$ quenching model, this can also clearly be seen in the orange contours of Fig.~\ref{fig:shmr_join}.  

Tab.~\ref{tab:M_scatter} compares the typical halo mass for a fixed stellar mass interval for each model, measured at the distribution peak along with $68$ per cent interval. In general we find that for the stellar mass ranges evaluated, dwarfs in the high-$z$ quenching model tend to reside in more massive haloes on average compared to the reference model. The difference becomes more pronounced at lower stellar mass, peaking with $0.15 \dex$ difference for the lowest mass range. Additionally, when evaluating the SHMR using present day halo masses we should consider the possibility that our formulation for orphan mass loss \citep{OLeary2021} may play a roll in artificially broadening the distribution, if that formulation strips too aggressively. We find that in our reference model orphans contribute between $\sim13$ and $\sim30$ per cent of the dwarf population $f_{\mathrm{orph}}$, while the high-$z$ quenching model tends to produce a lower orphan fraction ranging between $\sim10$ and $\sim20$ per cent.

Some recent work \citep{Nadler2019, Nadler2020, Wang2021} suggests that the singular power-law relation can be extrapolated into the ultra-faint dwarf regime for satellites down to $\lmstar \approx 2$. However these models have only been constrained down to $\lmstar \approx 8$ and are unable to self consistently reproduce the observed star formation history of these systems (see Sec.~\ref{sec:dwarf_sfh}). We find the introduction of high-$z$ quenching substantially alters the SHMR at low masses. In particular we find these models produce an over-abundance of galaxies in the $\lmstar \in [3,5)$ largely residing in haloes with $M^{\mathrm{peak}}_{\mathrm{h}} \lesssim M_{\mathrm{q}}$. This is an obvious consequence of our model which quenches star formation in these haloes but does not disrupt their constituent galaxy, effectively preventing their growth to higher masses along the standard SHMR. Without observables to validate such trends \textit{the SHMR should not be extrapolated beyond its range of constraint.}

\begin{table*}
	\centering
	\caption[Typical halo peak mass values for fixed stellar mass intervals]{Typical halo peak ($M^{\mathrm{peak}}_{\mathrm{h}}$) and current ($M^{z=0}_{\mathrm{h}}$ ) mass for a range of stellar mass intervals in each model. Masses are measured at the distribution peak with $68$ per cent interval measured at iso-density levels. $N_{\mathrm{gal}}$ shows the number of galaxies in each mass band and $f_{\mathrm{orph}}$ indicates the fraction of those galaxies that are orphans. All mass values are expressed in $\log_{10}(m/M_{\odot})$ units}
	\label{tab:M_scatter}
    \begin{tabular}{lcccccccr} 
		\hline
        \multirow{2}{*}{$m_*$} & \multicolumn{4}{c}{Reference} & \multicolumn{4}{c}{High-$z$ quench} \\
                                & $N_{\mathrm{gal}}$      & $f_{\mathrm{orph}}$ &  $M^{\mathrm{peak}}_{\mathrm{h}}$   &   $M^{z=0}_{\mathrm{h}}$   & $N_{\mathrm{gal}}$     & $f_{\mathrm{orph}}$   &   $M^{\mathrm{peak}}_{\mathrm{h}}$   &   $M^{z=0}_{\mathrm{h}}$   \\
		\hline
		\hline
        $\in [5.0, 5.5)$ & $133160$ & $0.31$ & $9.44_{-0.82}^{+0.32}$ & $9.55_{-2.07}^{+0.28}$ & $9950$ & $0.20$ & $9.58_{-0.14}^{+0.23}$ & $9.78_{-1.31}^{+0.22}$ \\
        $\in [5.5, 6.0)$ & $65613$ & $0.26$ & $9.77_{-0.79}^{+0.28}$ & $9.85_{-2.03}^{+0.28}$ & $12248$ & $0.14$ & $9.96_{-0.24}^{+0.13}$ & $10.01_{-0.98}^{+0.22}$ \\
        $\in [6.0, 6.5)$ & $32929$ & $0.22$ & $10.09_{-0.74}^{+0.24}$ & $10.13_{-1.93}^{+0.29}$ & $10701$ & $0.12$ & $10.20_{-0.26}^{+0.12}$ & $10.21_{-0.90}^{+0.22}$ \\
        $\in [6.5, 7.0)$ & $16459$ & $0.21$ & $10.39_{-0.66}^{+0.19}$ & $10.40_{-1.75}^{+0.30}$ & $8033$ & $0.12$ & $10.41_{-0.26}^{+0.13}$ & $10.42_{-0.94}^{+0.23}$ \\
        $\in [7.0, 7.5)$ & $8473$ & $0.19$ & $10.63_{-0.54}^{+0.18}$ & $10.62_{-1.63}^{+0.31}$ & $5626$ & $0.12$ & $10.61_{-0.23}^{+0.13}$ & $10.61_{-0.82}^{+0.25}$ \\
        $\in [7.5, 8.0)$ & $4515$ & $0.19$ & $10.85_{-0.42}^{+0.16}$ & $10.84_{-1.41}^{+0.34}$ & $3688$ & $0.13$ & $10.81_{-0.20}^{+0.13}$ & $10.80_{-0.71}^{+0.27}$ \\
        $\in [8.0, 8.5)$ & $2428$ & $0.17$ & $11.05_{-0.27}^{+0.15}$ & $11.03_{-1.03}^{+0.37}$ & $2398$ & $0.12$ & $10.99_{-0.17}^{+0.13}$ & $10.98_{-0.62}^{+0.29}$ \\
        $\in [8.5, 9.0)$ & $1496$ & $0.13$ & $11.23_{-0.18}^{+0.14}$ & $11.22_{-0.68}^{+0.35}$ & $1619$ & $0.10$ & $11.20_{-0.16}^{+0.10}$ & $11.18_{-0.44}^{+0.26}$ \\
		\hline
	\end{tabular}
\end{table*}

\subsection{Satellite populations}
Although our models have been fit assuming a global mass function, the majority of observed low mass systems are satellite galaxies of either the Milky Way or Andromeda. In this section we provide a comparison between the locally observed satellite mass functions and radial distributions compared with simulated systems of similar mass. Our simulated systems are all isolated galaxies (type 0) and selected based on both stellar and halo mass. When searching for simulated Milky Way and Andromeda analogues we use the stellar and halo mass ranges noted in Tab.~\ref{tab:obsmass}.
\begin{table}
	\centering
	\caption[Observed Milky Way and Andromeda properties]{Observational estimates of Milky Way and Andromeda  stellar and halo masses. Milky Way stellar and halo mass estimates are from \citet{BlandHawthorn2016}. Andromeda stellar and halo mass estimates are from \citet{Sick2015} and \citet{Diaz2014} respectively.}
	\label{tab:obsmass}
    \begin{tabular}{lccr} 
		\hline
		Name & $m_*$  & $M_{\mathrm{h}}^{\mathrm{obs}}$ & $M_{\mathrm{h}}^{\mathrm{sim}}$ \\
		\hline
		\hline
        Milky Way & $10.70^{+0.08}_{-0.10}$ & $12.11^{+0.09}_{-0.11}$ & $12.43^{+0.27}_{-0.31}$ \\
        Andromeda & $11.01^{+0.09}_{-0.08}$ & $12.23^{+0.07}_{-0.08}$ & $12.86^{+0.27}_{-0.29}$  \\
		\hline
	\end{tabular}
\end{table}

\begin{figure}
	\includegraphics[width=\columnwidth]{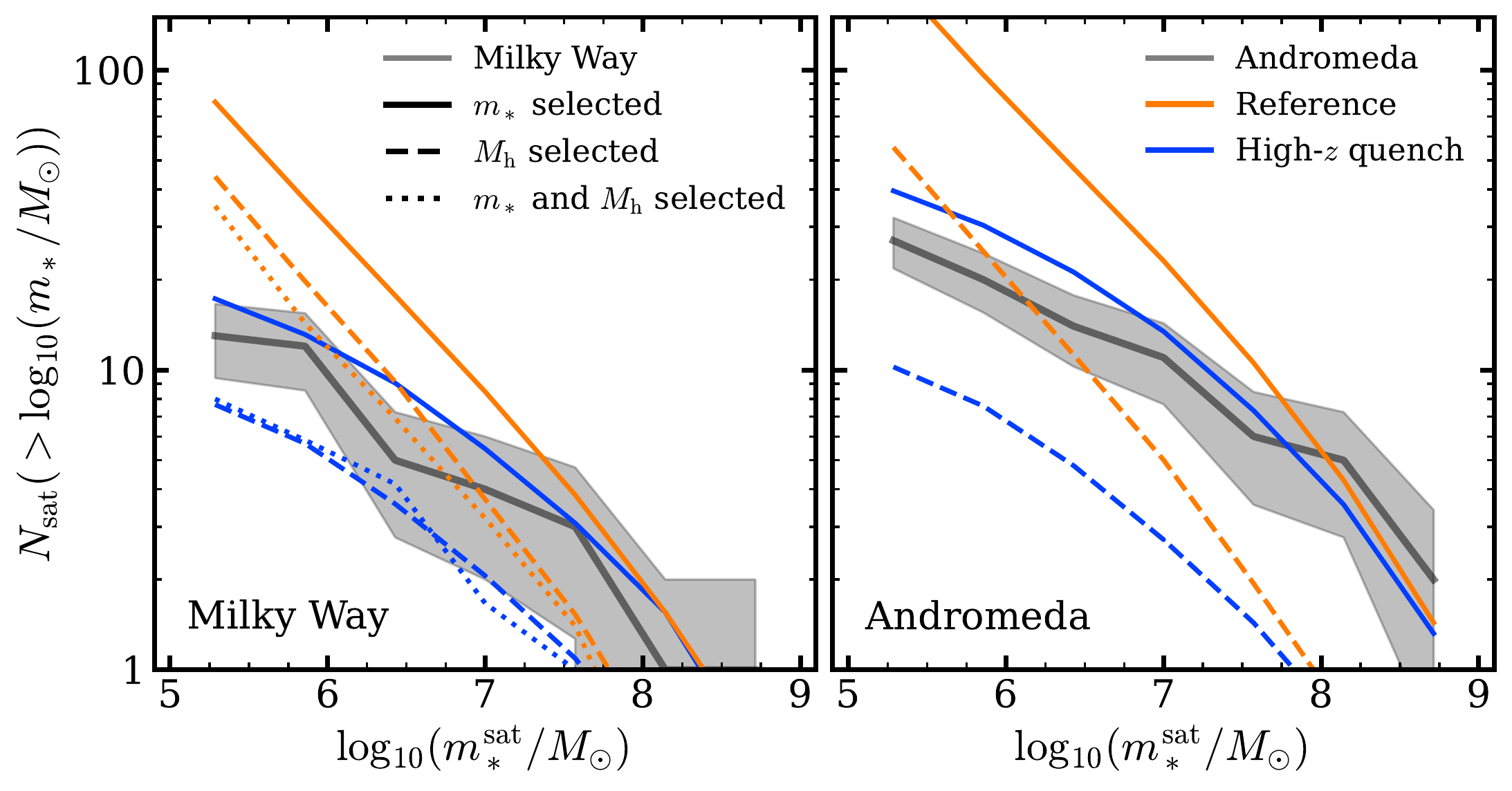} 
    \caption[Cumulative satellite stellar mass function around Milky Way and Andromeda like hosts]{Cumulative satellite stellar mass function for satellites with $5\leq \lmstar <9$ around Milky Way (left panel) and Andromeda (right panel) like hosts. The grey line and shaded region shows the observed stellar mass functions with Poisson error about the number count in each mass bin. Line colour indicates whether the mass function was generated from the reference model (orange) or the high-$z$ quenching model (blue). Solid lines show the mass function when selecting hosts based on stellar mass, dashed lines by halo mass and dotted lines by both stellar and halo mass.}
	\label{fig:sat_smf}
\end{figure}
Fig.~\ref{fig:sat_smf} shows the cumulative satellite mass function for the Milky Way (left panel) and Andromeda (right panel) under three different host selection criteria. In the first case we select simulated hosts based on stellar mass only (solid lines). Here we find that the reference model significantly over predicts the number of satellites for both Milky Way and Andromeda analogues. Enabling high-$z$ quenching substantially improves agreement with observation, exhibiting only a mild over prediction of abundances for satellites $\lmstar \lesssim 7.0$ around Andromeda analogues. In the second case we select simulated hosts based on halo mass (dashed lines). With respect to the Milky Way the reference model is consistent with observation for $\lmstar \gtrsim 7.0$ with a significant over prediction of satellite abundances at lower masses. Although halo mass selection does improve overall the  normalisation compared with observations there is a much steeper slope which drives the over abundances at the lowest masses. Meanwhile, the high-$z$ quenching model reproduces the general trend of the observed SMF but tends to undercut at all masses compared to the observed data. In the case of comparison to Andromeda we find that selecting hosts based on halo mass under predicts the abundances of satellites in both the reference case and high-$z$ quenching case. Finally, we check the case where we select hosts based on both stellar mass and halo mass. For the Milky Way we find that selecting based on both mass measures produces a mass function very similar to the halo mass only case. For Andromeda like systems we are unable to locate any simulated systems that meet both the stellar mass \textit{and} halo mass estimates from observation. In general, our simulated analogues (for both the Milky Way and Andromeda) reside in more massive haloes on average. Tab.~\ref{tab:obsmass} shows the average simulated halo masses where we can see there is no overlap in the observed and simulated halo ranges for Andromeda analogues. The implications of this mismatch are discussed in Sec.~\ref{sec:normalcy}.

\begin{figure}
	\includegraphics[width=\columnwidth]{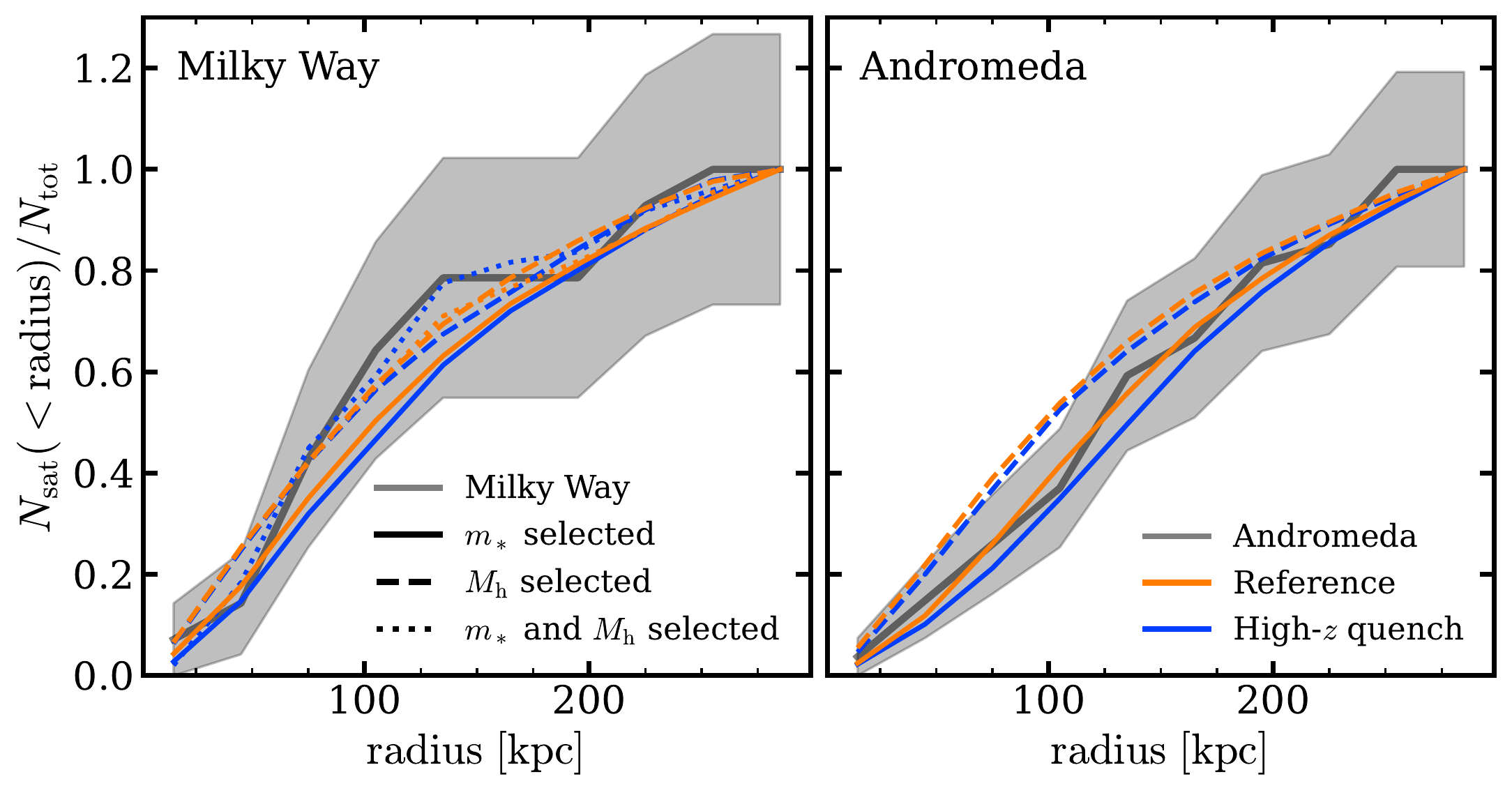}
    \caption[Normalised cumulative radial distribution of satellites around Milky Way and Andromeda like hosts]{Normalised cumulative radial distribution of satellites with $5\leq \lmstar <9$ within $300\kpc$ of Milky Way (left panel) and Andromeda (right panel) like hosts. The grey line and shaded region shows the observed radial distribution of satellites with Poisson error about the number count in each mass bin. Line colour indicates whether the sample was generated from the reference model (orange) or the logistic model (blue). Solid lines show the distribution when selecting hosts based on stellar mass, dashed lines by halo mass and dotted lines by both stellar and halo mass. Each line is normalised by the total number of satellites identified in the specified mass and radial ranges.}
	\label{fig:sat_dist}
\end{figure}
Fig.~\ref{fig:sat_dist} illustrates the cumulative satellite distribution around Milky Way (left panel) and Andromeda (right panel) like hosts. Here we find that when selecting hosts based on stellar mass both the reference and high-$z$ quenching model produce a satellite distribution in agreement with observation. For Milky Way analogues even selecting based on halo mass continues to produce a satellite distribution in line with observed trends. For Andromeda analogues we find that selecting based on halo mass tends to produce a slightly more centrally concentrated satellite distribution. This could of course be driven by the large mismatch in the simulated vs observed halo masses for these analogues.

\subsection{Star formation history}\label{sec:dwarf_sfh}
In this section we evaluate the star formation history (SFH) of low mass systems in this model. We trace back the main branches of the $z=0$ galaxy population and compute the average SFH, and stellar mass build up in discrete mass bins.

\begin{figure*}
	\includegraphics[width=\textwidth,height=\textheight,keepaspectratio]{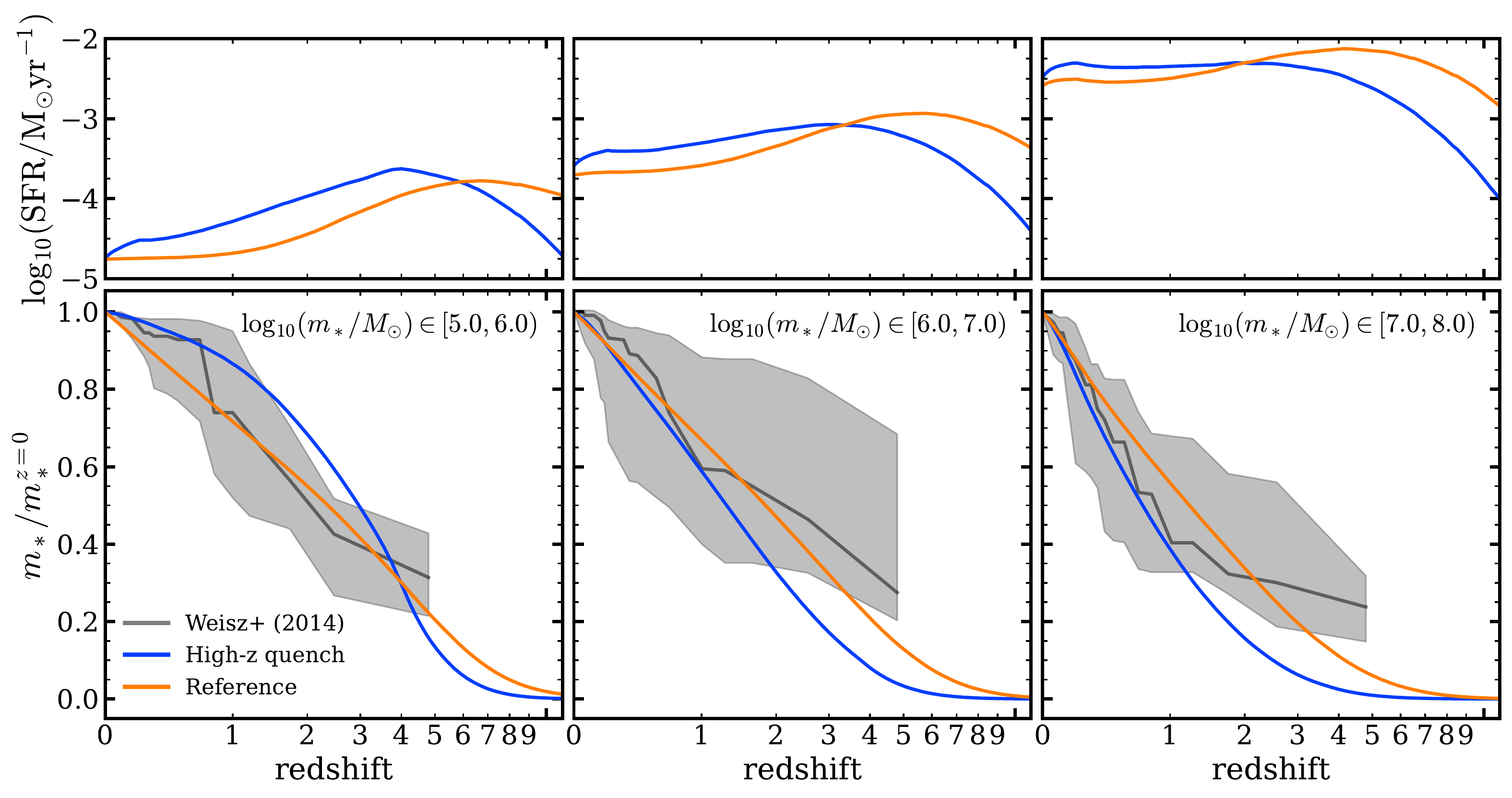}
    \caption[Star formation histories in the dwarf regime]{Star formation history in galaxies over time under the reference model (orange panels) and high-$z$ quenching model (blue panels) in three mass intervals. Upper panels show the average star formation rate as a function of redshift. Lower panels show the average cumulative mass growth for the same sample of galaxies. Grey lines illustrate the observed star formation history and $68$ per cent confidence interval for local dwarfs as computed by \citet{Weisz2014}}
	\label{fig:sfh}
\end{figure*}
The top row panel of Fig.~\ref{fig:sfh} compares the star formation history of the reference \emerge model (orange lines) with that of our high-$z$ quenching model (blue lines). Both models share some qualitative similarities. In both model variants we see that less massive systems experience peak SFR at higher redshift. For equivalent mass ranges the reference model experiences peak SFR earlier than the high-$z$ quenching model. For the two most massive bins the reference case also exhibits a higher peak SFR, for the lowest mass range the high-$z$ quenching model has a higher peak SFR. Further, we can see that in the case of high-$z$ quenching, peak SFR tends to occur very near to the redshift where $M^{\mathrm{peak}}_{\mathrm{q}}$ is reached. 

The bottom panels on Fig.~\ref{fig:sfh} illustrates the cumulative star formation history as a fraction of the $z=0$ stellar mass for the same sample of simulated galaxies as in the top panel. Although lowest mass systems in the reference model experience their peak SFR at higher redshift than the high-$z$ quenching model, we find that low mass systems in the high-$z$ quenching model are built up more rapidly than in the standard model. However, we find that both models are largely consistent with the observed average SFH determined by \citet{Weisz2014} (grey lines and shaded region).

\begin{figure}
	\includegraphics[width=\columnwidth]{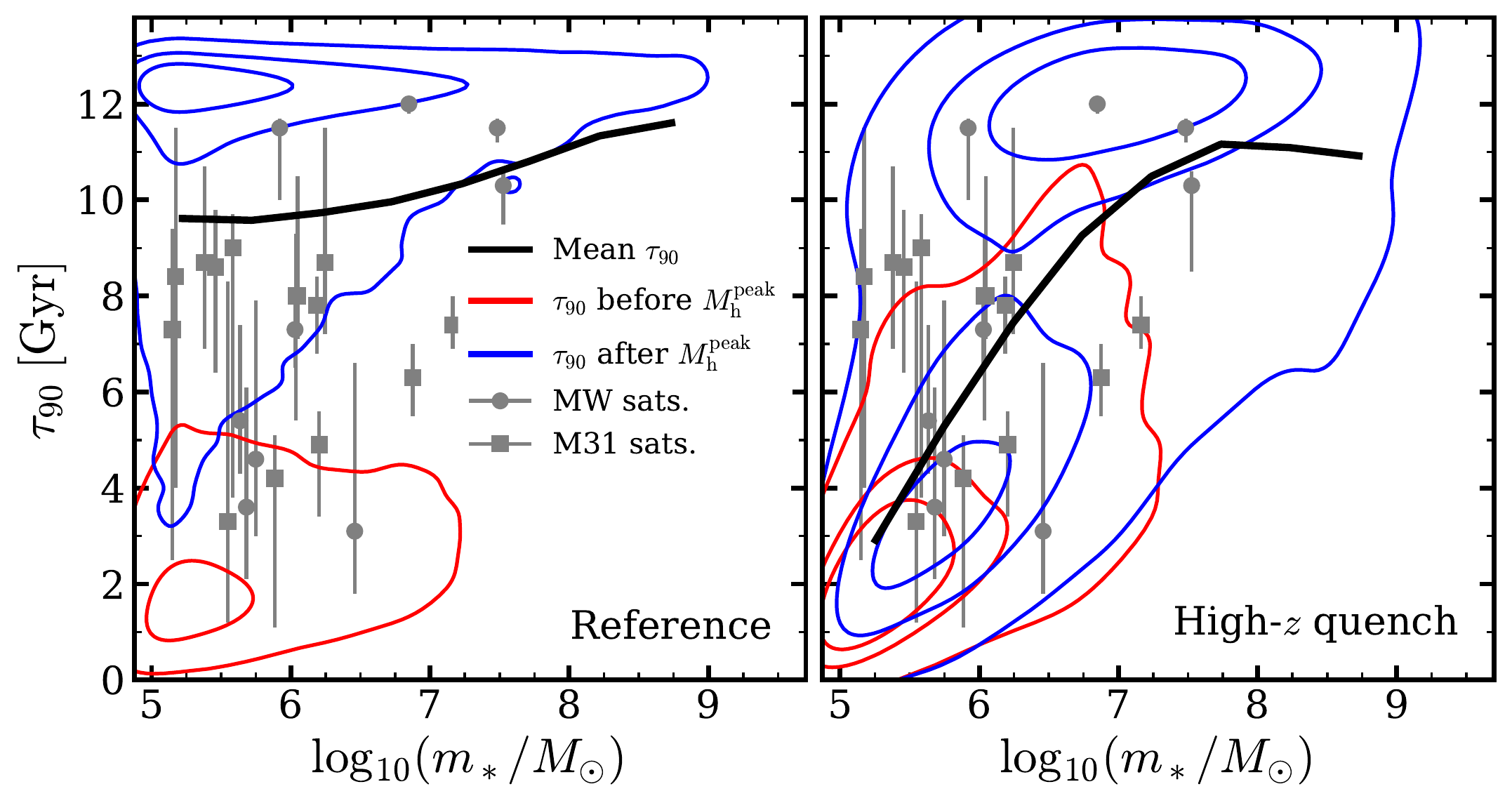}
    \caption[The $90$ per cent formation timescale in the dwarf regime]{The $90$ per cent formation timescale $\tau_{90}$, red contours indicated galaxies where the formation timescale occurred before halo peak mass. Blue contours show systems where formation occurred after peak halo mass. Black lines show the average formation timescale for all systems. Grey points show the formation timescale for individual observed dwarf satellites (see table~\ref{tab:obsdat})}
	\label{fig:t90}
\end{figure}
Another, perhaps more revealing measure of galaxy growth is the formation time. As a galaxy does not form in an instant we instead attribute the formation time as the $90$ per cent formation timescale $\tau_{90}$. This value specifies the cosmological time when a galaxy reached $90$ per cent of its present day mass. Fig.~\ref{fig:t90} compares the formation time of Milky Way and Andromeda satellites (grey points), with the satellite populations around our simulated analogues. In the first panel we can see that the reference case in general produces a mean formation time that is too long compared with observations, while also failing to reproduce the diversity of formation times seen in the data. The coloured lines indicated the iso-proportion contours for the formation time of galaxies in our simulations. Particularly in the case of the reference model we can see that \emerge produces a strongly bi-modal distribution of formation times at low masses. As previously stated, galaxy mass can be directly linked with its associated peak halo mass $M_h^{\mathrm{peak}}$. In Fig.\ref{fig:t90} we separate the galaxy sample to distinguish whether a galaxy formed before or after $M_h^{\mathrm{peak}}$. The blue contours indicate systems that formed after the halo reached $M_h^{\mathrm{peak}}$ and the red contours show galaxies that formed prior to $M_h^{\mathrm{peak}}$. The reference case shows that a majority of low mass systems are formed at late times after halo peak mass indicating that these objects experience significant stellar growth, coasting on their remaining cold gas reservoir, even while their host halo is no longer gaining mass. When high-$z$ quenching is enabled we can see that although the majority of galaxies still reach $\tau_{90}$ after peak mass, a significant portion of these systems form at much earlier times with a mean formation time in much better agreement with observed data. Furthermore, we can see that enabling high-$z$ quenching better reproduces the diversity of formation times seen in local satellites. That these low mass galaxies primarily form after $M_h^{\mathrm{peak}}$ but at early times, indicates that the majority of these systems have their star formation shut off during their coasting phase as opposed to active buildup.

\begin{figure}
	\includegraphics[width=\columnwidth]{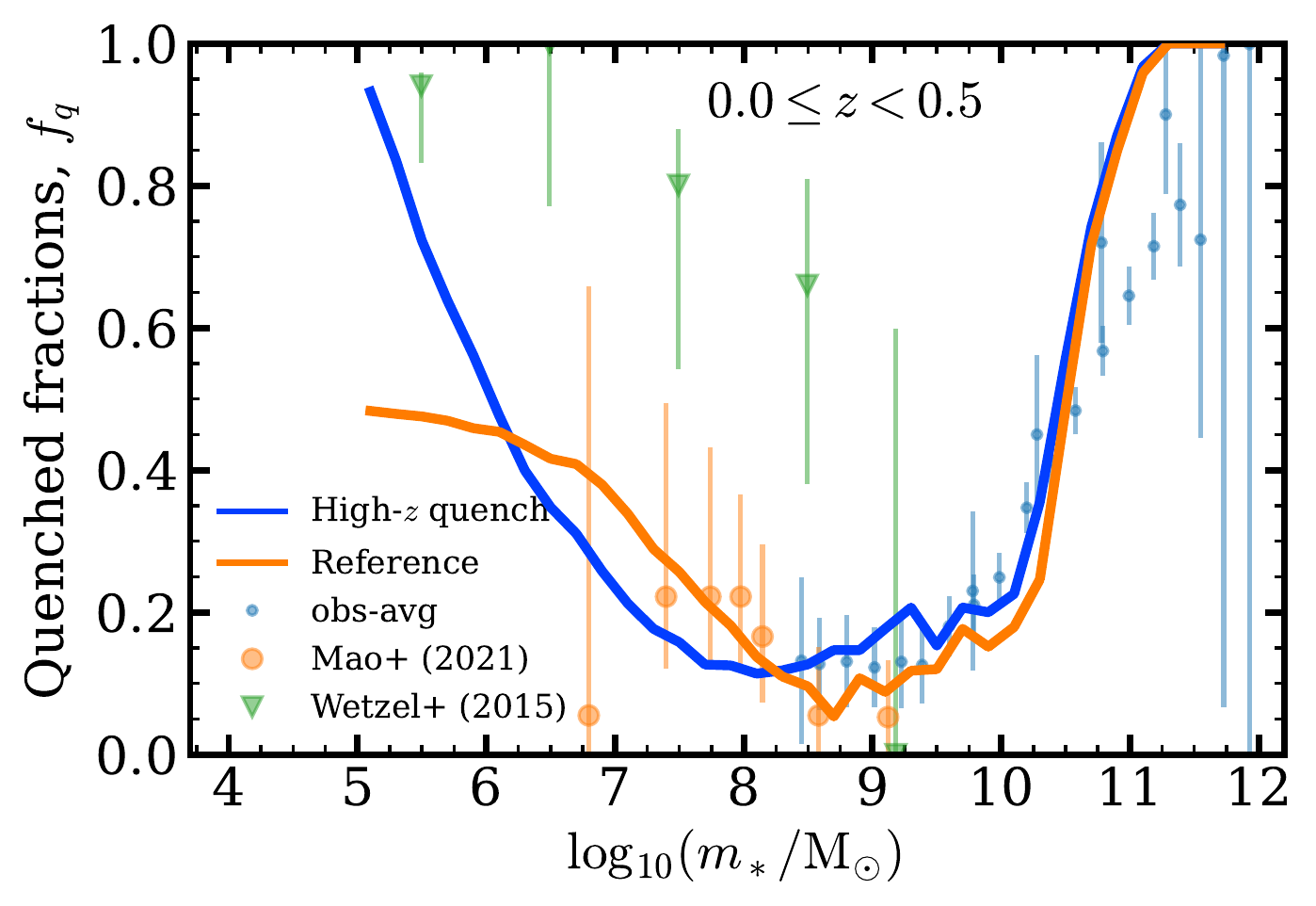}
    \caption[Galaxy quenched fractions in the dwarf regime]{Quenched fractions as a function of mass at $z=0$ for our reference (orange line) and high-$z$ quenching model (blue line). Observed quenched fractions in the dwarf regime are indicated by orange circles and green triangles. The orange circles indicate results from the SAGA survey \citep{Geha2017, Mao2021}, and the green triangles show quenched fractions for Milky Way and Andromeda satellites \citep{Wetzel2015}. At higher masses the observed average quenched fractions are indicated by blue points. A more complete view of the observable range at higher masses is illustrated in Fig.~\ref{fig:app_fq}.}
	\label{fig:fq}
\end{figure}
The consequences of these early formation times and premature quenching should therefore be observable in the present day star forming properties of our simulated galaxies. Fig.~\ref{fig:fq} shows the $z=0$ quenched fractions as a function of mass between our two model options. Although quenched fractions are not constrained for $\lmstar < 8.5$ there are still observed trends at low masses that we should consider. Unsurprisingly, both models exhibit nearly identical quenched fractions in the range used to constrain the model. For dwarf galaxies we see that the reference model possess a higher quenched fraction than the high-$z$ quenching model for $6 \lesssim \lmstar \lesssim 8.0$. Toward lower masses the reference model starts to saturate at near 50 per cent quenched. Conversely, the high-$z$ quenching model shows a rapidly increasing quenched fraction toward low masses, nearing 100 per cent quenched by $\lmstar = 5$.

\section{Discussion and Conclusions}\label{sec:discussion}
\subsection{Halo mass threshold for galaxy formation}
Our model suggests that there should be very few galaxies $\lmstar\gtrsim 5$ with $\lpmvir \lesssim 9 $ which is in agreement with other theoretical models that show high-$z$ quenching becomes important for haloes $\lpmvir \lesssim 9-10$ \citep{Thoul1996, Bullock2000, Somerville2002, Kuhlen2013, Sawala2015, Nadler2019}. Our model differs from other empirical models that include high-$z$ quenching in that we do not enforce that all haloes below some threshold should remain dark. The quenching mechanism we implement suppresses star formation as a function of halo mass and time, effectively penalising star formation in late forming haloes. Recent hydrodynamical models indicate that this distinction may be necessary in order to reproduce the properties of the ultra-faint dwarf (UFD) galaxies with $\lmstar \lesssim 5$ \citep{GarrisonKimmel2019, Munshi2021, Applebaum2021}. Additionally, using high resolution hydrodynamical simulations \citet{Schauer2021} suggest that a minimum halo mass between $10^6 \sim 10^7\Msun$ is required to begin star formation, depending on free streaming velocities and the strength of the Lyman-Werner Background. With respect to our work, that would indicate that whatever UFDs exist were likely strongly impacted by a high-$z$ quenching process and reside in haloes with a halo peak mass spanning only $\sim2$ orders of magnitude.

\subsection{Is the Local Group representative?}\label{sec:normalcy}
\begin{figure}
	\includegraphics[width=\columnwidth]{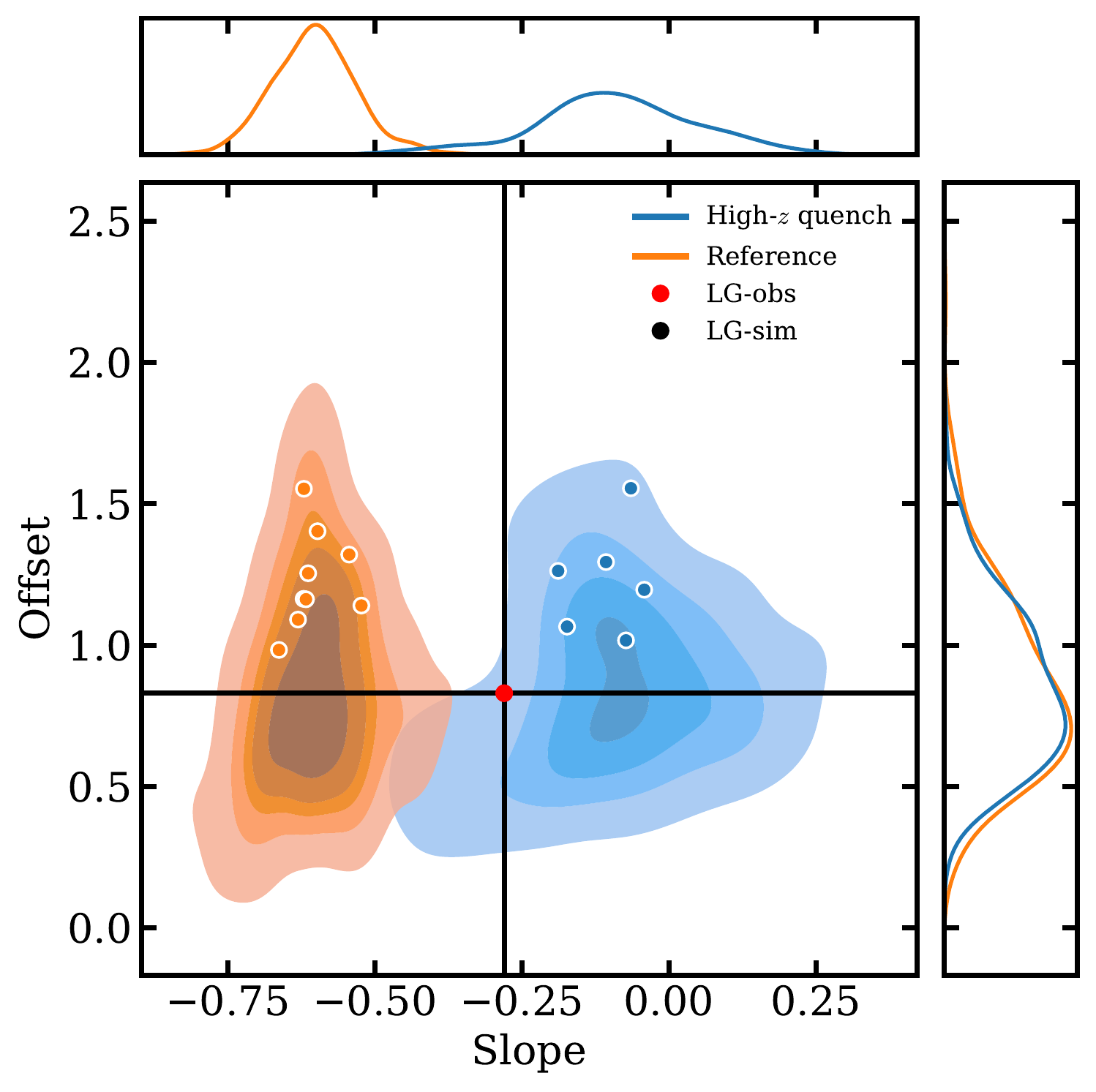}
    \caption[Locally measured mass function slope and offset from the average around simulated Milky Way like hosts]{The relation between the locally measured density offset from the global averaged and the locally observed power law slope of the mass function for Milky Way like systems. Coloured contours show the distribution of these two properties based on the $2\Mpc$ spherical volume around simulated systems similar to the Milky Way, based only on stellar mass selection. Solid points indicate Milky Way analogues that also host a massive Andromeda like companion within $1\Mpc$. The red marker shows the locally measured slope and offset based on the galaxies shown in Table~\ref{tab:obsdat} and illustrated by the observed mass function in Fig.~\ref{fig:GSMF}.}
	\label{fig:norm}
\end{figure}

The work presented so far has assumed that the Local Group is somewhat average with a local mass function that is representative of larger volumes. However we have so far not provided any quantitative analysis of how `normal' the Local Group is. Fig.~\ref{fig:norm} shows how the distribution in local SMF slope relates to the vertical offset (in\dex) from the global average for stellar mass selected Milky Way analogues. The local slope is defined by the power-law index measured in a $2\Mpc$ sphere around each analogue. Solid points indicate Milky Way like systems that \textit{also} host one Andromeda like companion within $1\Mpc$. The locally measured slope and offset derived from the data in table~\ref{tab:obsdat} is indicated by the red point. If our reference model (orange contours/points) is correct it would indicated that our locally observed SMF slope is a substantial outlier compared to simulated Milky Way analogues. However, the locally observed offset from the global SMF is average with respect to simulated systems. Additionally, we can see that the simulated systems that host an Andromeda companion are more likely to reside closer to the global slope than those without.
What is possibly more interesting is that even after fitting our model to Local Group data (blue contours/points) we find that the local SMF still does not posses an average local SMF slope compared with simulated systems, but is still within the simulated range. In this case the observed slope is steeper than the majority of simulated analogues. These deviations could indicate that assuming a global slope from local data is not adequate to constrain dwarf systems. To complicate the matter, halo masses for our simulated analogues tend to be higher than estimates for both Milky Way or Andromeda. The mismatch may or could indicate that the Local Group is indeed less dense than comparable counterparts and be subject to an accretion history that is not representative for haloes with comparable mass. 

Finally, we have assumed that adopting a spherical volume is sufficient to describe the Local Group. \citet{Putman2021} instead propose a prolate Local Group surface to more accurately define group membership. It is unclear how our simplification might impact our goodness of fit or alter the demographics of Local Group analogues. This would be an interesting area for future work. Additionally, we have assumed that currently confirmed dwarf observations provide sample completeness down to $\lmstar\sim5$ \citep{Tollerud2008}. Other recent works suggest that the current sample of galaxies is incomplete due to selection bias and the local observed mass function ought to be higher \citep{Loveday2015,  Jethwa2018, Newton2018, Nadler2019, Nadler2020, DrlicaWagner2020}, which could also explain the dip in the SMF for $\lmstar \lesssim 6.0$ (see Fig.~\ref{fig:GSMF}). In testing we found that allowing for a steeper mass function at low masses can negate the need for a high-$z$ quenching model and tend to prefer an earlier quenching with a lower $M_{\mathrm{q}}$.

\subsection{Other model options}
The foundation for this work is the introduction of a basic model for high-$z$ quenching that has been implemented in order to match the characteristics of locally observed dwarf systems. This high-$z$ quenching model only shows that we can reproduce locally observed populations by imposing restrictions on star formation at high redshift in a manner consistent with expectations from reionizaton. Limitations in our own implementation might restrict our ability to use this model to place direct constraints on the epoch of reionization. As implemented this model does not and cannot accurately capture the richness of evolution imposed by a reionization process. In particular, all models we tested suppress galaxy formation via an instantaneous quenching mechanism although there is evidence to suggest star formation may be periodic or even continue long after reionization \citep{Geha2012, Skillman2017}. The need for stars to form beyond reionization may explain why our models tend to favour such a late quenching scale $a_{\mathrm{q}}$. 

A key tenet of the empirical approach is to select the most simple model required to reproduce the observed data only increasing the complexity when the data demands it. For the purposes of the analysis in this work we have instead adopted the slightly more complicated model in favour of its flexibility to constrain with the inclusion of future observables. There are however several sensible model options that could be explored if more information on the abundances and star formation histories of dwarfs becomes available.

\subsubsection{Augmented baryon conversion efficiency}
Another reasonable approach to this problem would be to modify the standard baryon conversion efficiency with an additional term to further suppress the SFR in low mass haloes.
\begin{equation}\label{eq:low_eff2}
    \epsilon(M) = 2\epsilon_{\mathrm{N}} \left[\left(\frac{M}{M_{\mathrm{q}}}\right)^{-\alpha} + \left(\frac{M}{M_1}\right)^{-\beta} + \left(\frac{M}{M_1}\right)^{\gamma}\right]^{-1}
\end{equation}
While this approach is sensible and fits well within the current paradigm of the model, in testing we found that SMF data alone was insufficient to appropriately constrain the larger number of free parameters of this variant. In particular the redshift evolution of the terms $\alpha$ and $M_{\mathrm{q}}$ is difficult to determine without more detailed measurements on the star formation history of dwarfs. While the other models we test can quench and prevent star formation in haloes entirely, this method does not allow for empty haloes and does not introduce any explicit quenching mechanism.

\subsubsection{Quenching timescales}
Instead of shutting down star formation instantaneously we could instead set a timer similar to the standard quenching treatment in \emerge, allowing a more prolonged period for star formation after `reionization'. This would likely result in lower values for $M_{\mathrm{q}}$ and $a_{\mathrm{q}}$. This option was not explored but it may bring our high-$z$ quenching models in closer alignment with other models for reionization in cosmological simulations.

\subsubsection{Disk destruction}
As noted in \citet{OLeary2021} one of the lacking features of this model is the absence of proximity merging. Fig.~\ref{fig:sat_dist} does not indicate that we have excessive satellites at small radii, at least not in the mass ranges we explore. However, \citet{GarrisonKimmel2017} and \citet{Sawala2017} indicate that the presence of disk potential is significant for depletion of satellites in the inner halo. In our model the suppression of the SMF is accomplished between the baryon conversion efficiency, the high-$z$ quenching and through tidal stripping. In reality this is likely a more complex process that includes contributions from mergers or disruption due to the extended physical size of a host system. Here we have not explored the relative contributions of each of our model options on the abundance on dwarfs, but including additional mechanisms may alter our best fit model parameters. However, the parameters needed for these various mechanisms may be degenerate and locating observations that can independently constrain each parameter is problematic.

\subsection{Further analysis}\label{sec:dwarf_futurework}
Beyond the exploration of the SHMR there are some additional studies that could be performed to better understand the lives and ultimate fate of galaxies in low mass haloes.

The bulk of this work has focused on the global statistics of dwarf galaxies. A deeper study of local satellite dwarfs may provide additional validation as these objects are most readily observable. In particular, a better understanding of when the Milky Way and Andromeda acquired their satellites may help relate the distribution and star formation histories of these systems. Understanding these correlations could help verify our predictions or constrain the model.

One direct consequence of our model is the existence of dark haloes. Other works explore this topic \citep{Sawala2015, Sawala2016, Fitts2018} and establish an occupation fraction around $\lesssim 10$ per cent for $\lpmvir\lesssim 9$. The observational confirmation of dark haloes would provide vital information for breaking parameter degeneracies, more accurately determining the impact of reionization physics on the SMHR and the broader relationship between galaxies and haloes. Further work in this area with \emerge should include studies of dark haloes and their properties.

Furthermore, our models for high-$z$ quenching are rather \textit{ad hoc} and may not accurately represent how this process should evolve with time. An alternative approach would be to utilise machine learning to a inform a more appropriate parameterisation. \textsc{GalaxyNet} \citep{Moster2021} has already shown success for developing more accurate empirical prescriptions (see Sec.~\ref{sec:emerge_updates}). Future improvements may make it possible to do the same with the high-$z$ quenching model. The primary developments needed would be for the neural network to process complete  merger trees instead of isolated snapshots, this is necessary as the dwarf population is constrained by a low redshift observable that is driven by a high redshift process.

Finally, for this work our ability to fit the model with our chosen simulation was severely restricted by computational limitations of \emerge. At present \emerge, and other modern tree-based galaxy formation codes, parallelise by distributing the input halo merger trees across cpu cores such that each individual tree is in shared memory and accessed by only a single task. For simulations in moderate volumes of moderate resolution this is a completely adequate approach. When scaling these codes to more highly resolved simulations there are often a handful of very massive trees that dominate the computation time leaving all other tasks idle while the massive trees are processed. This places a hard limitation on the rate that parameter space can be explored and by extension our ability to quantitatively compare models variations. Future studies of dwarf galaxy formation in \emerge would benefit massively from refactoring the code to perform a more advance branch based parallelism that can allow for a more even distribution of the workload across many cores.

\subsection{Conclusions}

Over the course of this work we have developed an empirical model for galaxy formation with the goal of placing constraints on the stellar mass-halo mass relationship for galaxies down to $\lmstar = 5$. We achieve this goal by introducing an empirical quenching mechanism that approximates the effects of a reionization process. This model is constrained by our extension of the global galaxy stellar mass function using the mass function observed in the Local Group volume out to $2\Mpc$.

While our model has been fit assuming a global mass function, we show that our simulated galaxy sample reproduces both the number density and radial distributions of satellites for the Milky Way and Andromeda. Additionally, our simulated galaxies are able to reproduce properties of observed systems beyond stellar mass. In particular we find that for the $z=0$ sample, nearly 100 per cent of galaxies are quenched at $\lmstar\sim 5$. This is $50$ per cent higher than a model that does not include high-$z$ quenching. Further, we show that introducing high-$z$ quenching produces a population of galaxies that better matches the star formation histories and formation timescales of observes dwarf satellites.

We show that the SHMR can be extended as a power-law down to at least $\lpmvir \approx 10$. Models including high-$z$ quenching indicated substantially reduced scatter in peak halo mass at a fixed stellar mass and show that the elimination of high scatter galaxies galaxies in the tail of the distribution are the primary mechanism that flattens the SMF at low masses. Further inspection of the SHMR shows that there should be almost no galaxies down to $\lmstar = 5$ with a peak halo mass lower than $\lpmvir \lesssim 9.0$. This is consistent with our model implementations which predict quenching in haloes with $\lpmvir \lesssim 9.3$ by $z=4$ (see table~\ref{tab:mod_fit})

Incorporating a model which suppresses star formation in low mass-high redshift haloes produces a population of galaxies that is consistent with observables in both the total number density and in the bulk star formation history of galaxies. Our results indicate that as few as two additional free parameters are necessary to reproduce the number density of dwarf galaxies. However, we propose a more complex $3$ parameter logistic model as the framework for model variations due to its adaptability. 

\section*{Acknowledgements}

We thank all authors who provide their data in electronic form. The cosmological simulations used in this work were partially carried out at the Odin Cluster at the Max Planck Computing and Data Facility and SuperMUC-NG in Garchging. The authors gratefully acknowledge the Gauss Centre for Supercomputing e.V. (www.gauss-centre.eu) for funding this project by providing computing time on the GCS Supercomputer SuperMUC-NG at Leibniz Supercomputing Centre (www.lrz.de) via the project number pn72bu. The authors acknowldege the computing time provided by the cluster ``rusty'' of the Simons Foundation in New York City. 
UPS is supported by a Flatiron Research Fellowship (FRF) at the Center of Computational Astrophysics at the Flatiron Institute. The Flatiron Institute is supported by the Simons Foundation.
Finally, we thank the developers of \texttt{Astropy} \citep{AstropyCollaboration2013,AstropyCollaboration2018}, \texttt{NumPy} \citep{vanderWalt2011}, \texttt{SciPy} \citep{Virtanen2020}, \texttt{Jupyter} \citep{RaganKelley2014}, \texttt{Matplotlib} \citep{Hunter2007}, for their very useful free software. The Astrophysics Data Service (ADS) and \texttt{arXiv} preprint repository were used extensively in this work.

\section*{Data Availability}

The data will be made available based on reasonable request to the corresponding author.



\bibliographystyle{mnras}
\bibliography{references} 

\begin{thebibliography}{}
\makeatletter
\relax
\def\mn@urlcharsother{\let\do\@makeother \do\$\do\&\do\#\do\^\do\_\do\%\do\~}
\def\mn@doi{\begingroup\mn@urlcharsother \@ifnextchar [ {\mn@doi@}
  {\mn@doi@[]}}
\def\mn@doi@[#1]#2{\def\@tempa{#1}\ifx\@tempa\@empty \href
  {http://dx.doi.org/#2} {doi:#2}\else \href {http://dx.doi.org/#2} {#1}\fi
  \endgroup}
\def\mn@eprint#1#2{\mn@eprint@#1:#2::\@nil}
\def\mn@eprint@arXiv#1{\href {http://arxiv.org/abs/#1} {{\tt arXiv:#1}}}
\def\mn@eprint@dblp#1{\href {http://dblp.uni-trier.de/rec/bibtex/#1.xml}
  {dblp:#1}}
\def\mn@eprint@#1:#2:#3:#4\@nil{\def\@tempa {#1}\def\@tempb {#2}\def\@tempc
  {#3}\ifx \@tempc \@empty \let \@tempc \@tempb \let \@tempb \@tempa \fi \ifx
  \@tempb \@empty \def\@tempb {arXiv}\fi \@ifundefined
  {mn@eprint@\@tempb}{\@tempb:\@tempc}{\expandafter \expandafter \csname
  mn@eprint@\@tempb\endcsname \expandafter{\@tempc}}}

\bibitem[\protect\citeauthoryear{{Applebaum}, {Brooks}, {Christensen},
  {Munshi}, {Quinn}, {Shen}  \& {Tremmel}}{{Applebaum}
  et~al.}{2021}]{Applebaum2021}
{Applebaum} E.,  {Brooks} A.~M.,  {Christensen} C.~R.,  {Munshi} F.,  {Quinn}
  T.~R.,  {Shen} S.,   {Tremmel} M.,  2021, \mn@doi [\apj]
  {10.3847/1538-4357/abcafa}, \href
  {https://ui.adsabs.harvard.edu/abs/2021ApJ...906...96A} {906, 96}

\bibitem[\protect\citeauthoryear{{Astropy Collaboration} et~al.,}{{Astropy
  Collaboration} et~al.}{2013}]{AstropyCollaboration2013}
{Astropy Collaboration} et~al., 2013, \mn@doi [\aap]
  {10.1051/0004-6361/201322068}, \href
  {https://ui.adsabs.harvard.edu/abs/2013A&A...558A..33A} {558, A33}

\bibitem[\protect\citeauthoryear{{Astropy Collaboration} et~al.,}{{Astropy
  Collaboration} et~al.}{2018}]{AstropyCollaboration2018}
{Astropy Collaboration} et~al., 2018, \mn@doi [\aj] {10.3847/1538-3881/aabc4f},
  \href {https://ui.adsabs.harvard.edu/abs/2018AJ....156..123A} {156, 123}

\bibitem[\protect\citeauthoryear{{Beck} et~al.,}{{Beck}
  et~al.}{2016}]{Beck2016}
{Beck} A.~M.,  et~al., 2016, \mn@doi [\mnras] {10.1093/mnras/stv2443}, \href
  {https://ui.adsabs.harvard.edu/abs/2016MNRAS.455.2110B} {455, 2110}

\bibitem[\protect\citeauthoryear{{Behroozi}, {Wechsler}  \& {Wu}}{{Behroozi}
  et~al.}{2013a}]{Behroozi2013}
{Behroozi} P.~S.,  {Wechsler} R.~H.,   {Wu} H.-Y.,  2013a, \mn@doi [\apj]
  {10.1088/0004-637X/762/2/109}, \href
  {http://adsabs.harvard.edu/abs/2013ApJ...762..109B} {762, 109}

\bibitem[\protect\citeauthoryear{{Behroozi}, {Wechsler}, {Wu}, {Busha},
  {Klypin}  \& {Primack}}{{Behroozi} et~al.}{2013b}]{Behroozi2013d}
{Behroozi} P.~S.,  {Wechsler} R.~H.,  {Wu} H.-Y.,  {Busha} M.~T.,  {Klypin}
  A.~A.,   {Primack} J.~R.,  2013b, \mn@doi [\apj]
  {10.1088/0004-637X/763/1/18}, \href
  {http://adsabs.harvard.edu/abs/2013ApJ...763...18B} {763, 18}

\bibitem[\protect\citeauthoryear{{Behroozi}, {Wechsler}, {Hearin}  \&
  {Conroy}}{{Behroozi} et~al.}{2019}]{Behroozi2019}
{Behroozi} P.,  {Wechsler} R.~H.,  {Hearin} A.~P.,   {Conroy} C.,  2019,
  \mn@doi [\mnras] {10.1093/mnras/stz1182}, \href
  {https://ui.adsabs.harvard.edu/abs/2019MNRAS.488.3143B} {488, 3143}

\bibitem[\protect\citeauthoryear{{Bell} \& {de Jong}}{{Bell} \& {de
  Jong}}{2001}]{Bell2001}
{Bell} E.~F.,  {de Jong} R.~S.,  2001, \mn@doi [\apj] {10.1086/319728}, \href
  {https://ui.adsabs.harvard.edu/abs/2001ApJ...550..212B} {550, 212}

\bibitem[\protect\citeauthoryear{{Benitez-Llambay} \&
  {Fumagalli}}{{Benitez-Llambay} \& {Fumagalli}}{2021}]{Benitez-Llambay2021}
{Benitez-Llambay} A.,  {Fumagalli} M.,  2021, \mn@doi [\apjl]
  {10.3847/2041-8213/ac3006}, \href
  {https://ui.adsabs.harvard.edu/abs/2021ApJ...921L...9B} {921, L9}

\bibitem[\protect\citeauthoryear{{Bland-Hawthorn} \&
  {Gerhard}}{{Bland-Hawthorn} \& {Gerhard}}{2016}]{BlandHawthorn2016}
{Bland-Hawthorn} J.,  {Gerhard} O.,  2016, \mn@doi [\araa]
  {10.1146/annurev-astro-081915-023441}, \href
  {https://ui.adsabs.harvard.edu/abs/2016ARA&A..54..529B} {54, 529}

\bibitem[\protect\citeauthoryear{{Bullock}, {Kravtsov}  \&
  {Weinberg}}{{Bullock} et~al.}{2000}]{Bullock2000}
{Bullock} J.~S.,  {Kravtsov} A.~V.,   {Weinberg} D.~H.,  2000, \mn@doi [\apj]
  {10.1086/309279}, \href
  {https://ui.adsabs.harvard.edu/abs/2000ApJ...539..517B} {539, 517}

\bibitem[\protect\citeauthoryear{{Carlsten}, {Greene}, {Peter}, {Beaton}  \&
  {Greco}}{{Carlsten} et~al.}{2020}]{Carlsten2020}
{Carlsten} S.~G.,  {Greene} J.~E.,  {Peter} A. H.~G.,  {Beaton} R.~L.,
  {Greco} J.~P.,  2020, arXiv e-prints, \href
  {https://ui.adsabs.harvard.edu/abs/2020arXiv200602443C} {p. arXiv:2006.02443}

\bibitem[\protect\citeauthoryear{{Cole}, {Weisz}, {Dolphin}, {Skillman},
  {McConnachie}, {Brooks}  \& {Leaman}}{{Cole} et~al.}{2014}]{Cole2014}
{Cole} A.~A.,  {Weisz} D.~R.,  {Dolphin} A.~E.,  {Skillman} E.~D.,
  {McConnachie} A.~W.,  {Brooks} A.~M.,   {Leaman} R.,  2014, \mn@doi [\apj]
  {10.1088/0004-637X/795/1/54}, \href
  {https://ui.adsabs.harvard.edu/abs/2014ApJ...795...54C} {795, 54}

\bibitem[\protect\citeauthoryear{{Cooray}}{{Cooray}}{2006}]{Cooray2006}
{Cooray} A.,  2006, \mn@doi [\mnras] {10.1111/j.1365-2966.2005.09747.x}, \href
  {https://ui.adsabs.harvard.edu/abs/2006MNRAS.365..842C} {365, 842}

\bibitem[\protect\citeauthoryear{{Diaz}, {Koposov}, {Irwin}, {Belokurov}  \&
  {Evans}}{{Diaz} et~al.}{2014}]{Diaz2014}
{Diaz} J.~D.,  {Koposov} S.~E.,  {Irwin} M.,  {Belokurov} V.,   {Evans} N.~W.,
  2014, \mn@doi [\mnras] {10.1093/mnras/stu1210}, \href
  {https://ui.adsabs.harvard.edu/abs/2014MNRAS.443.1688D} {443, 1688}

\bibitem[\protect\citeauthoryear{{Drlica-Wagner} et~al.,}{{Drlica-Wagner}
  et~al.}{2020}]{DrlicaWagner2020}
{Drlica-Wagner} A.,  et~al., 2020, \mn@doi [\apj] {10.3847/1538-4357/ab7eb9},
  \href {https://ui.adsabs.harvard.edu/abs/2020ApJ...893...47D} {893, 47}

\bibitem[\protect\citeauthoryear{{Efstathiou}, {Bond}  \& {White}}{{Efstathiou}
  et~al.}{1992}]{Efstathiou1992}
{Efstathiou} G.,  {Bond} J.~R.,   {White} S.~D.~M.,  1992, \mn@doi [\mnras]
  {10.1093/mnras/258.1.1P}, \href
  {https://ui.adsabs.harvard.edu/abs/1992MNRAS.258P...1E} {258, 1P}

\bibitem[\protect\citeauthoryear{{Fattahi}, {Navarro}  \& {Frenk}}{{Fattahi}
  et~al.}{2020}]{Fattahi2020}
{Fattahi} A.,  {Navarro} J.~F.,   {Frenk} C.~S.,  2020, \mn@doi [\mnras]
  {10.1093/mnras/staa375}, \href
  {https://ui.adsabs.harvard.edu/abs/2020MNRAS.493.2596F} {493, 2596}

\bibitem[\protect\citeauthoryear{{Fitts} et~al.,}{{Fitts}
  et~al.}{2018}]{Fitts2018}
{Fitts} A.,  et~al., 2018, \mn@doi [\mnras] {10.1093/mnras/sty1488}, \href
  {https://ui.adsabs.harvard.edu/abs/2018MNRAS.479..319F} {479, 319}

\bibitem[\protect\citeauthoryear{{Gallart} et~al.,}{{Gallart}
  et~al.}{2015}]{Gallart2015}
{Gallart} C.,  et~al., 2015, \mn@doi [\apjl] {10.1088/2041-8205/811/2/L18},
  \href {https://ui.adsabs.harvard.edu/abs/2015ApJ...811L..18G} {811, L18}

\bibitem[\protect\citeauthoryear{{Garrison-Kimmel} et~al.,}{{Garrison-Kimmel}
  et~al.}{2017}]{GarrisonKimmel2017}
{Garrison-Kimmel} S.,  et~al., 2017, \mn@doi [\mnras] {10.1093/mnras/stx1710},
  \href {https://ui.adsabs.harvard.edu/abs/2017MNRAS.471.1709G} {471, 1709}

\bibitem[\protect\citeauthoryear{{Garrison-Kimmel} et~al.,}{{Garrison-Kimmel}
  et~al.}{2019}]{GarrisonKimmel2019}
{Garrison-Kimmel} S.,  et~al., 2019, \mn@doi [\mnras] {10.1093/mnras/stz1317},
  \href {https://ui.adsabs.harvard.edu/abs/2019MNRAS.487.1380G} {487, 1380}

\bibitem[\protect\citeauthoryear{{Geha}, {Blanton}, {Yan}  \& {Tinker}}{{Geha}
  et~al.}{2012}]{Geha2012}
{Geha} M.,  {Blanton} M.~R.,  {Yan} R.,   {Tinker} J.~L.,  2012, \mn@doi [\apj]
  {10.1088/0004-637X/757/1/85}, \href
  {https://ui.adsabs.harvard.edu/abs/2012ApJ...757...85G} {757, 85}

\bibitem[\protect\citeauthoryear{{Geha} et~al.,}{{Geha}
  et~al.}{2017}]{Geha2017}
{Geha} M.,  et~al., 2017, \mn@doi [\apj] {10.3847/1538-4357/aa8626}, \href
  {https://ui.adsabs.harvard.edu/abs/2017ApJ...847....4G} {847, 4}

\bibitem[\protect\citeauthoryear{{Hahn} \& {Abel}}{{Hahn} \&
  {Abel}}{2011}]{Hahn2011}
{Hahn} O.,  {Abel} T.,  2011, \mn@doi [\mnras]
  {10.1111/j.1365-2966.2011.18820.x}, \href
  {http://adsabs.harvard.edu/abs/2011MNRAS.415.2101H} {415, 2101}

\bibitem[\protect\citeauthoryear{{Hirano}, {Hosokawa}, {Yoshida}, {Omukai}  \&
  {Yorke}}{{Hirano} et~al.}{2015}]{Hirano2015}
{Hirano} S.,  {Hosokawa} T.,  {Yoshida} N.,  {Omukai} K.,   {Yorke} H.~W.,
  2015, \mn@doi [\mnras] {10.1093/mnras/stv044}, \href
  {https://ui.adsabs.harvard.edu/abs/2015MNRAS.448..568H} {448, 568}

\bibitem[\protect\citeauthoryear{{Hunter}}{{Hunter}}{2007}]{Hunter2007}
{Hunter} J.~D.,  2007, \mn@doi [Computing in Science and Engineering]
  {10.1109/MCSE.2007.55}, \href
  {https://ui.adsabs.harvard.edu/abs/2007CSE.....9...90H} {9, 90}

\bibitem[\protect\citeauthoryear{{Jethwa}, {Erkal}  \& {Belokurov}}{{Jethwa}
  et~al.}{2018}]{Jethwa2018}
{Jethwa} P.,  {Erkal} D.,   {Belokurov} V.,  2018, \mn@doi [\mnras]
  {10.1093/mnras/stx2330}, \href
  {https://ui.adsabs.harvard.edu/abs/2018MNRAS.473.2060J} {473, 2060}

\bibitem[\protect\citeauthoryear{{Klypin}, {Kravtsov}, {Valenzuela}  \&
  {Prada}}{{Klypin} et~al.}{1999}]{Klypin1999}
{Klypin} A.,  {Kravtsov} A.~V.,  {Valenzuela} O.,   {Prada} F.,  1999, \mn@doi
  [\apj] {10.1086/307643}, \href
  {https://ui.adsabs.harvard.edu/abs/1999ApJ...522...82K} {522, 82}

\bibitem[\protect\citeauthoryear{{Kravtsov} \& {Manwadkar}}{{Kravtsov} \&
  {Manwadkar}}{2022}]{Kravtsov2022}
{Kravtsov} A.,  {Manwadkar} V.,  2022, \mn@doi [\mnras]
  {10.1093/mnras/stac1439}, \href
  {https://ui.adsabs.harvard.edu/abs/2022MNRAS.514.2667K} {514, 2667}

\bibitem[\protect\citeauthoryear{{Kuhlen}, {Madau}  \& {Krumholz}}{{Kuhlen}
  et~al.}{2013}]{Kuhlen2013}
{Kuhlen} M.,  {Madau} P.,   {Krumholz} M.~R.,  2013, \mn@doi [\apj]
  {10.1088/0004-637X/776/1/34}, \href
  {https://ui.adsabs.harvard.edu/abs/2013ApJ...776...34K} {776, 34}

\bibitem[\protect\citeauthoryear{{Kulkarni}, {Visbal}  \& {Bryan}}{{Kulkarni}
  et~al.}{2021}]{Kulkarni2021}
{Kulkarni} M.,  {Visbal} E.,   {Bryan} G.~L.,  2021, \mn@doi [\apj]
  {10.3847/1538-4357/ac08a3}, \href
  {https://ui.adsabs.harvard.edu/abs/2021ApJ...917...40K} {917, 40}

\bibitem[\protect\citeauthoryear{{Lewis}, {Challinor}  \& {Lasenby}}{{Lewis}
  et~al.}{2000}]{Lewis2000}
{Lewis} A.,  {Challinor} A.,   {Lasenby} A.,  2000, \mn@doi [\apj]
  {10.1086/309179}, \href {http://adsabs.harvard.edu/abs/2000ApJ...538..473L}
  {538, 473}

\bibitem[\protect\citeauthoryear{{Loveday} et~al.,}{{Loveday}
  et~al.}{2015}]{Loveday2015}
{Loveday} J.,  et~al., 2015, \mn@doi [\mnras] {10.1093/mnras/stv1013}, \href
  {https://ui.adsabs.harvard.edu/abs/2015MNRAS.451.1540L} {451, 1540}

\bibitem[\protect\citeauthoryear{{Mao}, {Geha}, {Wechsler}, {Weiner},
  {Tollerud}, {Nadler}  \& {Kallivayalil}}{{Mao} et~al.}{2021}]{Mao2021}
{Mao} Y.-Y.,  {Geha} M.,  {Wechsler} R.~H.,  {Weiner} B.,  {Tollerud} E.~J.,
  {Nadler} E.~O.,   {Kallivayalil} N.,  2021, \mn@doi [\apj]
  {10.3847/1538-4357/abce58}, \href
  {https://ui.adsabs.harvard.edu/abs/2021ApJ...907...85M} {907, 85}

\bibitem[\protect\citeauthoryear{{Martin}, {de Jong}  \& {Rix}}{{Martin}
  et~al.}{2008}]{Martin2008}
{Martin} N.~F.,  {de Jong} J. T.~A.,   {Rix} H.-W.,  2008, \mn@doi [\apj]
  {10.1086/590336}, \href
  {https://ui.adsabs.harvard.edu/abs/2008ApJ...684.1075M} {684, 1075}

\bibitem[\protect\citeauthoryear{{McConnachie}}{{McConnachie}}{2012}]{McConnachie2012}
{McConnachie} A.~W.,  2012, \mn@doi [\aj] {10.1088/0004-6256/144/1/4}, \href
  {https://ui.adsabs.harvard.edu/abs/2012AJ....144....4M} {144, 4}

\bibitem[\protect\citeauthoryear{{Moore}, {Ghigna}, {Governato}, {Lake},
  {Quinn}, {Stadel}  \& {Tozzi}}{{Moore} et~al.}{1999}]{Moore1999}
{Moore} B.,  {Ghigna} S.,  {Governato} F.,  {Lake} G.,  {Quinn} T.,  {Stadel}
  J.,   {Tozzi} P.,  1999, \mn@doi [\apjl] {10.1086/312287}, \href
  {https://ui.adsabs.harvard.edu/abs/1999ApJ...524L..19M} {524, L19}

\bibitem[\protect\citeauthoryear{{Moster}, {Somerville}, {Maulbetsch}, {van den
  Bosch}, {Macci{\`o}}, {Naab}  \& {Oser}}{{Moster} et~al.}{2010}]{Moster2010}
{Moster} B.~P.,  {Somerville} R.~S.,  {Maulbetsch} C.,  {van den Bosch} F.~C.,
  {Macci{\`o}} A.~V.,  {Naab} T.,   {Oser} L.,  2010, \mn@doi [\apj]
  {10.1088/0004-637X/710/2/903}, \href
  {https://ui.adsabs.harvard.edu/abs/2010ApJ...710..903M} {710, 903}

\bibitem[\protect\citeauthoryear{{Moster}, {Naab}  \& {White}}{{Moster}
  et~al.}{2018}]{Moster2018}
{Moster} B.~P.,  {Naab} T.,   {White} S. D.~M.,  2018, \mn@doi [\mnras]
  {10.1093/mnras/sty655}, \href
  {https://ui.adsabs.harvard.edu/abs/2018MNRAS.477.1822M} {477, 1822}

\bibitem[\protect\citeauthoryear{{Moster}, {Naab}, {Lindstr{\"o}m}  \&
  {O'Leary}}{{Moster} et~al.}{2021}]{Moster2021}
{Moster} B.~P.,  {Naab} T.,  {Lindstr{\"o}m} M.,   {O'Leary} J.~A.,  2021,
  \mn@doi [\mnras] {10.1093/mnras/stab1449}, \href
  {https://ui.adsabs.harvard.edu/abs/2021MNRAS.507.2115M} {507, 2115}

\bibitem[\protect\citeauthoryear{{Munshi}, {Brooks}, {Applebaum},
  {Christensen}, {Sligh}  \& {Quinn}}{{Munshi} et~al.}{2021}]{Munshi2021}
{Munshi} F.,  {Brooks} A.,  {Applebaum} E.,  {Christensen} C.,  {Sligh} J.~P.,
   {Quinn} T.,  2021, arXiv e-prints, \href
  {https://ui.adsabs.harvard.edu/abs/2021arXiv210105822M} {p. arXiv:2101.05822}

\bibitem[\protect\citeauthoryear{{Nadler}, {Mao}, {Green}  \&
  {Wechsler}}{{Nadler} et~al.}{2019}]{Nadler2019}
{Nadler} E.~O.,  {Mao} Y.-Y.,  {Green} G.~M.,   {Wechsler} R.~H.,  2019,
  \mn@doi [\apj] {10.3847/1538-4357/ab040e}, \href
  {https://ui.adsabs.harvard.edu/abs/2019ApJ...873...34N} {873, 34}

\bibitem[\protect\citeauthoryear{{Nadler} et~al.,}{{Nadler}
  et~al.}{2020}]{Nadler2020}
{Nadler} E.~O.,  et~al., 2020, \mn@doi [\apj] {10.3847/1538-4357/ab846a}, \href
  {https://ui.adsabs.harvard.edu/abs/2020ApJ...893...48N} {893, 48}

\bibitem[\protect\citeauthoryear{{Newton}, {Cautun}, {Jenkins}, {Frenk}  \&
  {Helly}}{{Newton} et~al.}{2018}]{Newton2018}
{Newton} O.,  {Cautun} M.,  {Jenkins} A.,  {Frenk} C.~S.,   {Helly} J.~C.,
  2018, \mn@doi [\mnras] {10.1093/mnras/sty1085}, \href
  {https://ui.adsabs.harvard.edu/abs/2018MNRAS.479.2853N} {479, 2853}

\bibitem[\protect\citeauthoryear{{O'Leary}, {Moster}, {Naab}  \&
  {Somerville}}{{O'Leary} et~al.}{2021}]{OLeary2021}
{O'Leary} J.~A.,  {Moster} B.~P.,  {Naab} T.,   {Somerville} R.~S.,  2021,
  \mn@doi [\mnras] {10.1093/mnras/staa3746}, \href
  {https://ui.adsabs.harvard.edu/abs/2021MNRAS.501.3215O} {501, 3215}

\bibitem[\protect\citeauthoryear{{Planck Collaboration}}{{Planck
  Collaboration}}{2016}]{PlanckCollaboration2016}
{Planck Collaboration} 2016, \mn@doi [\aap] {10.1051/0004-6361/201525830},
  \href {http://adsabs.harvard.edu/abs/2016A%26A...594A..13P} {594, A13}

\bibitem[\protect\citeauthoryear{{Putman}, {Zheng}, {Price-Whelan}, {Grcevich},
  {Johnson}, {Tollerud}  \& {Peek}}{{Putman} et~al.}{2021}]{Putman2021}
{Putman} M.~E.,  {Zheng} Y.,  {Price-Whelan} A.~M.,  {Grcevich} J.,  {Johnson}
  A.~C.,  {Tollerud} E.,   {Peek} J. E.~G.,  2021, \mn@doi [\apj]
  {10.3847/1538-4357/abe391}, \href
  {https://ui.adsabs.harvard.edu/abs/2021ApJ...913...53P} {913, 53}

\bibitem[\protect\citeauthoryear{{Ragan-Kelley}, {Perez}, {Granger}, {Kluyver},
  {Ivanov}, {Frederic}  \& {Bussonnier}}{{Ragan-Kelley}
  et~al.}{2014}]{RaganKelley2014}
{Ragan-Kelley} M.,  {Perez} F.,  {Granger} B.,  {Kluyver} T.,  {Ivanov} P.,
  {Frederic} J.,   {Bussonnier} M.,  2014, in AGU Fall Meeting Abstracts. pp
  H44D--07

\bibitem[\protect\citeauthoryear{{Rey} \& {Starkenburg}}{{Rey} \&
  {Starkenburg}}{2022}]{Rey2022}
{Rey} M.~P.,  {Starkenburg} T.~K.,  2022, \mn@doi [\mnras]
  {10.1093/mnras/stab3709}, \href
  {https://ui.adsabs.harvard.edu/abs/2022MNRAS.510.4208R} {510, 4208}

\bibitem[\protect\citeauthoryear{{Rey}, {Pontzen}, {Agertz}, {Orkney}, {Read}
  \& {Rosdahl}}{{Rey} et~al.}{2020}]{Rey2020}
{Rey} M.~P.,  {Pontzen} A.,  {Agertz} O.,  {Orkney} M. D.~A.,  {Read} J.~I.,
  {Rosdahl} J.,  2020, \mn@doi [\mnras] {10.1093/mnras/staa1640}, \href
  {https://ui.adsabs.harvard.edu/abs/2020MNRAS.497.1508R} {497, 1508}

\bibitem[\protect\citeauthoryear{{Ricotti}, {Gnedin}  \& {Shull}}{{Ricotti}
  et~al.}{2008}]{Ricotti2008}
{Ricotti} M.,  {Gnedin} N.~Y.,   {Shull} J.~M.,  2008, \mn@doi [\apj]
  {10.1086/590901}, \href
  {https://ui.adsabs.harvard.edu/abs/2008ApJ...685...21R} {685, 21}

\bibitem[\protect\citeauthoryear{{Sawala} et~al.,}{{Sawala}
  et~al.}{2015}]{Sawala2015}
{Sawala} T.,  et~al., 2015, \mn@doi [\mnras] {10.1093/mnras/stu2753}, \href
  {https://ui.adsabs.harvard.edu/abs/2015MNRAS.448.2941S} {448, 2941}

\bibitem[\protect\citeauthoryear{{Sawala} et~al.,}{{Sawala}
  et~al.}{2016a}]{Sawala2016}
{Sawala} T.,  et~al., 2016a, \mn@doi [\mnras] {10.1093/mnras/stv2597}, \href
  {https://ui.adsabs.harvard.edu/abs/2016MNRAS.456...85S} {456, 85}

\bibitem[\protect\citeauthoryear{{Sawala} et~al.,}{{Sawala}
  et~al.}{2016b}]{Sawala2016b}
{Sawala} T.,  et~al., 2016b, \mn@doi [\mnras] {10.1093/mnras/stw145}, \href
  {https://ui.adsabs.harvard.edu/abs/2016MNRAS.457.1931S} {457, 1931}

\bibitem[\protect\citeauthoryear{{Sawala}, {Pihajoki}, {Johansson}, {Frenk},
  {Navarro}, {Oman}  \& {White}}{{Sawala} et~al.}{2017}]{Sawala2017}
{Sawala} T.,  {Pihajoki} P.,  {Johansson} P.~H.,  {Frenk} C.~S.,  {Navarro}
  J.~F.,  {Oman} K.~A.,   {White} S. D.~M.,  2017, \mn@doi [\mnras]
  {10.1093/mnras/stx360}, \href
  {https://ui.adsabs.harvard.edu/abs/2017MNRAS.467.4383S} {467, 4383}

\bibitem[\protect\citeauthoryear{{Schauer}, {Glover}, {Klessen}  \&
  {Clark}}{{Schauer} et~al.}{2021}]{Schauer2021}
{Schauer} A. T.~P.,  {Glover} S. C.~O.,  {Klessen} R.~S.,   {Clark} P.,  2021,
  \mn@doi [\mnras] {10.1093/mnras/stab1953}, \href
  {https://ui.adsabs.harvard.edu/abs/2021MNRAS.507.1775S} {507, 1775}

\bibitem[\protect\citeauthoryear{{Sick}, {Courteau}, {Cuillandre}, {Dalcanton},
  {de Jong}, {McDonald}, {Simard}  \& {Tully}}{{Sick} et~al.}{2015}]{Sick2015}
{Sick} J.,  {Courteau} S.,  {Cuillandre} J.-C.,  {Dalcanton} J.,  {de Jong} R.,
   {McDonald} M.,  {Simard} D.,   {Tully} R.~B.,  2015, in {Cappellari} M.,
  {Courteau} S.,  eds,  IAU Symposium Vol. 311, Galaxy Masses as Constraints of
  Formation Models. pp 82--85 (\mn@eprint {arXiv} {1410.0017}),
  \mn@doi{10.1017/S1743921315003440}

\bibitem[\protect\citeauthoryear{{Simon}}{{Simon}}{2018}]{Simon2018}
{Simon} J.~D.,  2018, \mn@doi [\apj] {10.3847/1538-4357/aacdfb}, \href
  {https://ui.adsabs.harvard.edu/abs/2018ApJ...863...89S} {863, 89}

\bibitem[\protect\citeauthoryear{{Skillman} et~al.,}{{Skillman}
  et~al.}{2017}]{Skillman2017}
{Skillman} E.~D.,  et~al., 2017, \mn@doi [\apj] {10.3847/1538-4357/aa60c5},
  \href {https://ui.adsabs.harvard.edu/abs/2017ApJ...837..102S} {837, 102}

\bibitem[\protect\citeauthoryear{{Smercina}, {Bell}, {Price}, {D'Souza},
  {Slater}, {Bailin}, {Monachesi}  \& {Nidever}}{{Smercina}
  et~al.}{2018}]{Smercina2018}
{Smercina} A.,  {Bell} E.~F.,  {Price} P.~A.,  {D'Souza} R.,  {Slater} C.~T.,
  {Bailin} J.,  {Monachesi} A.,   {Nidever} D.,  2018, \mn@doi [\apj]
  {10.3847/1538-4357/aad2d6}, \href
  {https://ui.adsabs.harvard.edu/abs/2018ApJ...863..152S} {863, 152}

\bibitem[\protect\citeauthoryear{{Somerville}}{{Somerville}}{2002}]{Somerville2002}
{Somerville} R.~S.,  2002, \mn@doi [\apjl] {10.1086/341444}, \href
  {https://ui.adsabs.harvard.edu/abs/2002ApJ...572L..23S} {572, L23}

\bibitem[\protect\citeauthoryear{{Springel}}{{Springel}}{2005}]{Springel2005b}
{Springel} V.,  2005, \mn@doi [\mnras] {10.1111/j.1365-2966.2005.09655.x},
  \href {http://adsabs.harvard.edu/abs/2005MNRAS.364.1105S} {364, 1105}

\bibitem[\protect\citeauthoryear{{Thoul} \& {Weinberg}}{{Thoul} \&
  {Weinberg}}{1996}]{Thoul1996}
{Thoul} A.~A.,  {Weinberg} D.~H.,  1996, \mn@doi [\apj] {10.1086/177446}, \href
  {https://ui.adsabs.harvard.edu/abs/1996ApJ...465..608T} {465, 608}

\bibitem[\protect\citeauthoryear{{Tollerud}, {Bullock}, {Strigari}  \&
  {Willman}}{{Tollerud} et~al.}{2008}]{Tollerud2008}
{Tollerud} E.~J.,  {Bullock} J.~S.,  {Strigari} L.~E.,   {Willman} B.,  2008,
  \mn@doi [\apj] {10.1086/592102}, \href
  {https://ui.adsabs.harvard.edu/abs/2008ApJ...688..277T} {688, 277}

\bibitem[\protect\citeauthoryear{{Virtanen} et~al.,}{{Virtanen}
  et~al.}{2020}]{Virtanen2020}
{Virtanen} P.,  et~al., 2020, \mn@doi [Nature Methods]
  {https://doi.org/10.1038/s41592-019-0686-2}, \href {https://rdcu.be/b08Wh}
  {17, 261}

\bibitem[\protect\citeauthoryear{{Wang}, {Nadler}, {Mao}, {Adhikari},
  {Wechsler}  \& {Behroozi}}{{Wang} et~al.}{2021}]{Wang2021}
{Wang} Y.,  {Nadler} E.~O.,  {Mao} Y.-Y.,  {Adhikari} S.,  {Wechsler} R.~H.,
  {Behroozi} P.,  2021, arXiv e-prints, \href
  {https://ui.adsabs.harvard.edu/abs/2021arXiv210211876W} {p. arXiv:2102.11876}

\bibitem[\protect\citeauthoryear{{Weisz}, {Dolphin}, {Skillman}, {Holtzman},
  {Gilbert}, {Dalcanton}  \& {Williams}}{{Weisz} et~al.}{2014}]{Weisz2014}
{Weisz} D.~R.,  {Dolphin} A.~E.,  {Skillman} E.~D.,  {Holtzman} J.,  {Gilbert}
  K.~M.,  {Dalcanton} J.~J.,   {Williams} B.~F.,  2014, \mn@doi [\apj]
  {10.1088/0004-637X/789/2/147}, \href
  {https://ui.adsabs.harvard.edu/abs/2014ApJ...789..147W} {789, 147}

\bibitem[\protect\citeauthoryear{{Weisz}, {Dolphin}, {Skillman}, {Holtzman},
  {Gilbert}, {Dalcanton}  \& {Williams}}{{Weisz} et~al.}{2015}]{Weisz2015}
{Weisz} D.~R.,  {Dolphin} A.~E.,  {Skillman} E.~D.,  {Holtzman} J.,  {Gilbert}
  K.~M.,  {Dalcanton} J.~J.,   {Williams} B.~F.,  2015, \mn@doi [\apj]
  {10.1088/0004-637X/804/2/136}, \href
  {https://ui.adsabs.harvard.edu/abs/2015ApJ...804..136W} {804, 136}

\bibitem[\protect\citeauthoryear{{Weisz} et~al.,}{{Weisz}
  et~al.}{2019}]{Weisz2019}
{Weisz} D.~R.,  et~al., 2019, \mn@doi [\apjl] {10.3847/2041-8213/ab4b52}, \href
  {https://ui.adsabs.harvard.edu/abs/2019ApJ...885L...8W} {885, L8}

\bibitem[\protect\citeauthoryear{{Wetzel}, {Tollerud}  \& {Weisz}}{{Wetzel}
  et~al.}{2015}]{Wetzel2015}
{Wetzel} A.~R.,  {Tollerud} E.~J.,   {Weisz} D.~R.,  2015, \mn@doi [\apjl]
  {10.1088/2041-8205/808/1/L27}, \href
  {https://ui.adsabs.harvard.edu/abs/2015ApJ...808L..27W} {808, L27}

\bibitem[\protect\citeauthoryear{{Woo}, {Courteau}  \& {Dekel}}{{Woo}
  et~al.}{2008}]{Woo2008}
{Woo} J.,  {Courteau} S.,   {Dekel} A.,  2008, \mn@doi [\mnras]
  {10.1111/j.1365-2966.2008.13770.x}, \href
  {https://ui.adsabs.harvard.edu/abs/2008MNRAS.390.1453W} {390, 1453}

\bibitem[\protect\citeauthoryear{{van der Walt}, {Colbert}  \&
  {Varoquaux}}{{van der Walt} et~al.}{2011}]{vanderWalt2011}
{van der Walt} S.,  {Colbert} S.~C.,   {Varoquaux} G.,  2011, \mn@doi [CiSE]
  {10.1109/MCSE.2011.37}, \href
  {https://ui.adsabs.harvard.edu/abs/2011CSE....13b..22V} {13, 22}

\makeatother
\end{thebibliography}


\appendix
\section{Observed dwarf data}
The table in this section tabulates the dwarf catalogue data that was used to extend our model constraints to $\lmstar \approx 5$. The table additionally includes details on the distance of each system to the Milky Way and Andromeda (M31), as well as $90$ per cent formation timescales used for the comparisons in Section \ref{sec:dwarf_sfh}.
\begin{table*}
	\centering
	\caption[Properties of Local Group dwarf galaxies]{Catalogue of Local Group dwarf galaxies and their properties used in this work. The reference column indicates the source for the $\tau_{90}$ measurements.}
	\label{tab:obsdat}
	\resizebox{0.75\textwidth}{!}{\begin{tabular}{lcccccr} 
		\hline
		Galaxy Name   & $\log_{10}(m^*/\Msun)$  & $D_{\odot}[\kpc]$ & $D_{\mathrm{MW}}[\kpc]$ & $D_{\mathrm{M}31}[\kpc]$ & $\tau_{90}$ [Gyr] & Reference\\
		\hline
		\hline
        Sagittariusd-Sph  & 7.526 & 27 & 19 & 787 & $10.3^{+0.33}_{-1.82}$ & \citet{Weisz2014}\\
        LMC              & 9.380 & 51 & 50 & 807 & - & - \\
        SMC              & 8.867 & 64 & 61 & 807 & - & - \\
        UrsaMinor        & 5.748 & 76 & 78 & 754 & $4.63^{+3.27}_{-1.60}$ & \citet{Weisz2014}\\
        Draco            & 5.681 & 82 & 82 & 748 & $3.55^{+2.50}_{-1.52}$ & \citet{Weisz2014}\\
        Sculptor         & 6.459 & 86 & 86 & 761 & $3.09^{+3.53}_{-1.29}$ & \citet{Weisz2014}\\
        Sextans(1)       & 5.709 & 95 & 98 & 841 & - & - \\
        Carina           & 5.920 & 106 & 108 & 838 & $11.46^{+0.07}_{-1.49}$ & \citet{Weisz2014} \\
        Crater 2          & 5.408 & 118 & 116 & 886 & - & - \\
        Antlia 2          & 5.748 & 132 & 133 & 889 & - & - \\
        Fornax           & 7.483 & 139 & 141 & 768 & $11.46^{+0.20}_{-0.27}$ & \citet{Weisz2014}\\
        Canes Venatici(1) & 5.635 & 211 & 211 & 856 & $5.38^{+2.01}_{-1.13}$ & \citet{Weisz2014} \\
        Leo 2             & 6.030 & 233 & 236 & 897 & $7.29^{+0.63}_{-0.75}$ & \citet{Weisz2014}\\
        Leo 1             & 6.848 & 254 & 258 & 918 & $12.02^{+0.06}_{-0.20}$ & \citet{Weisz2014}\\
        Leo T             & 5.350 & 409 & 414 & 982 & $12.12^{+0.12}_{-0.06}$  & \citet{Weisz2014} \\
        Phoenix          & 6.091 & 415 & 415 & 864 & $10.56^{+0.63}_{-1.03}$ & \citet{Gallart2015} \\
        NGC6822          & 8.204 & 459 & 452 & 894 & - & - \\
        Andromeda XVI     & 5.736 & 476 & 480 & 319 & $7.88^{+0.56}_{-0.49}$ & \citet{Skillman2017}\\
        Andromeda XXIV    & 5.173 & 600 & 605 & 204 & $8.4^{+3.1}_{-4.4}$ & \citet{Weisz2019} \\
        NGC185           & 8.037 & 617 & 621 & 184 & - & - \\
        Andromeda XV      & 5.885 & 625 & 630 & 175 & $4.24^{+0.87}_{-3.13}$ & \citet{Skillman2017} \\
        Andromeda II      & 7.163 & 652 & 656 & 181 & $7.39^{+0.60}_{-0.51}$ & \citet{Skillman2017} \\
        Andromeda XXVIII  & 5.526 & 661 & 661 & 365 & $6.13^{+0.28}_{-1.75}$ & \citet{Skillman2017} \\
        Andromeda X       & 5.149 & 670 & 674 & 130 & $7.3^{+2.1}_{-4.8}$ & \citet{Weisz2019} \\
        NGC147           & 7.997 & 676 & 680 & 139 & - & - \\
        Andromeda XXX     & 5.318 & 682 & 686 & 144 & - & - \\
        Andromeda XVII    & 5.547 & 728 & 732 & 66  & $3.3^{+5.0}_{-2.1}$ & \citet{Weisz2019} \\
        Andromeda XXIX    & 5.459 & 731 & 734 & 187 & $8.6^{+1.2}_{-2.2}$ & \citet{Weisz2019} \\
        Andromeda I       & 6.876 & 745 & 749 & 55 & $6.29^{+0.67}_{-0.84}$ & \citet{Skillman2017} \\
        Andromeda III     & 6.204 & 748 & 752 & 73 & $4.93^{+0.67}_{-1.47}$ & \citet{Skillman2017} \\
        IC1613           & 8.204 & 755 & 758 & 518 & $11.46^{+0.63}_{-1.03}$ & \citet{Gallart2015} \\
        Cetus            & 6.651 & 755 & 756 & 678 & $4.3^{+0.63}_{-1.03}$ & \citet{Gallart2015} \\
        Andromeda XXXI    & 6.817 & 759 & 760 & 262  & - & - \\
        Andromeda VII     & 7.408 & 762 & 765 & 217 & - & - \\
        Andromeda IX      & 5.380 & 766 & 770 & 39  & $8.7^{+2.0}_{-1.8}$ & \citet{Weisz2019} \\
        LGS3             & 6.186 & 769 & 773 & 268 & $7.82^{+0.63}_{-1.03}$ & \citet{Gallart2015} \\
        Andromeda XXIII   & 6.246 & 769 & 774 & 126  & $8.7^{+2.8}_{-1.5}$ & \citet{Weisz2019} \\
        Andromeda XXXIII  & 6.283 & 773 & 779 & 348 & - & - \\
        Andromeda V       & 5.952 & 773 & 777 & 109  & - & - \\
        Andromeda XXXII   & 7.037 & 776 & 780 & 140 & - & - \\
        Andromeda VI      & 6.723 & 783 & 785 & 268 & - & - \\
        Andromeda XIV     & 5.584 & 794 & 798 & 161 & $9.0^{+0.7}_{-5.2}$ & \citet{Weisz2019} \\
        IC10             & 8.139 & 794 & 798 & 252 & - & - \\
        Leo A             & 6.982 & 798 & 803 & 1197 & $12.55^{+0.63}_{-1.03}$ & \citet{Gallart2015} \\
        M32              & 8.709 & 805 & 809 & 27 & - & - \\
        Andromeda XXV     & 6.037 & 813 & 817 & 90  & $8.0^{+1.3}_{-2.6}$ & \citet{Weisz2019}  \\
        Andromeda XIX     & 5.723 & 820 & 824 & 115 & - & - \\
        NGC205           & 8.723 & 824 & 828 & 46 & - & - \\
        Andromeda XXI     & 6.049 & 828 & 831 & 135 & $8.0^{+2.5}_{-0.9}$ & \citet{Weisz2019} \\
        Andromeda XXVII   & 5.283 & 828 & 832 & 77 & - & - \\
        Tucana           & 5.952 & 887 & 883 & 1352 & $4.11^{+0.63}_{-1.03}$ & \citet{Gallart2015} \\
        Pegasus-dIrr      & 7.024 & 920 & 921 & 474 & - & - \\
        WLM              & 7.838 & 933 & 933 & 835 & - & - \\
        Sagittarius-dIrr  & 6.748 & 1067 & 1059 & 1354 & - & - \\
        Aquarius         & 6.408 & 1072 & 1066 & 1170 & $11.1^{+0.02}_{-0.04}$ & \citet{Cole2014} \\
        Andromeda XVIII   & 5.903 & 1213 & 1217 & 457  & $9.2^{+2.1}_{-1.7}$ & \citet{Weisz2019} \\
        Antlia B          & 6.003 & 1294 & 1296 & 1963 & - & - \\
        NGC3109          & 8.085 & 1300 & 1301 & 1984 & - & - \\
        Antlia           & 6.318 & 1349 & 1350 & 2036 & - & - \\
        UGC4879          & 7.123 & 1361 & 1367 & 1394 & - & - \\
        Sextans B         & 7.920 & 1426 & 1429 & 1940 & - & - \\
        Sextans A         & 7.848 & 1432 & 1435 & 2024 & - & - \\
        Leo P             & 5.795 & 1622 & 1625 & 2048 & - & - \\
        KKR25            & 6.505 & 1923 & 1922 & 1869 & - & - \\
        ESO410-G005      & 6.748 & 1923 & 1922 & 1861 & - & - \\
        IC5152           & 8.350 & 1950 & 1945 & 2209 & - & - \\
		\hline
	\end{tabular}}
\end{table*}
\section{Model variants}\label{sec:appB}
\subsection{Model fits to observables}
In this section we show how each model variant agrees with the observable constraints of each model. The observables indicated in the figures in this section correspond to the actual data used during fitting.
\begin{figure*}
	\includegraphics[width=\textwidth,keepaspectratio]{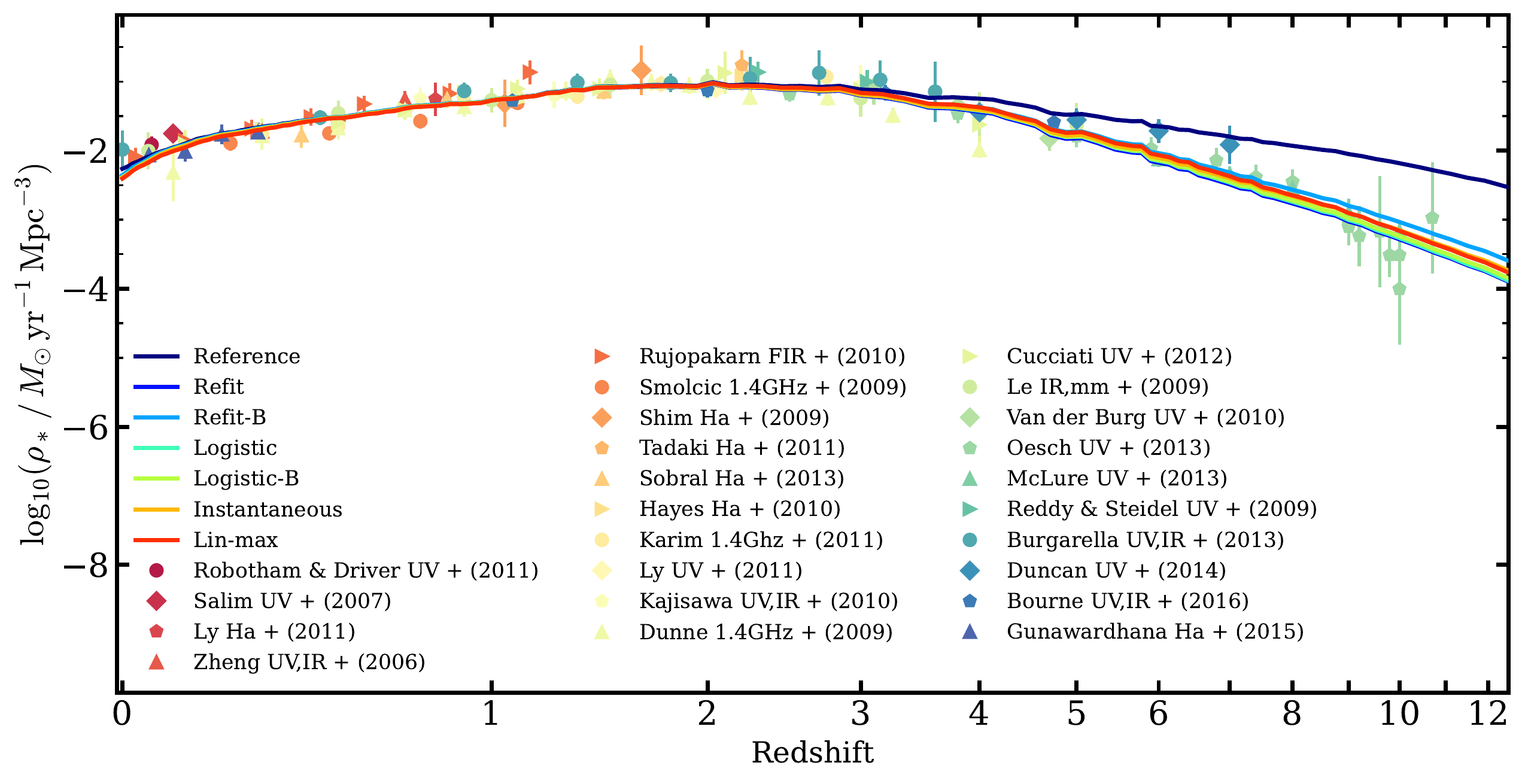}
	\centering
    \caption[]{The model predicted cosmic star formation rate density compared to the input observables for each of our model variants}
	\label{fig:app_csfrd}
\end{figure*}

\begin{figure*}
	\includegraphics[width=0.98\textwidth,keepaspectratio]{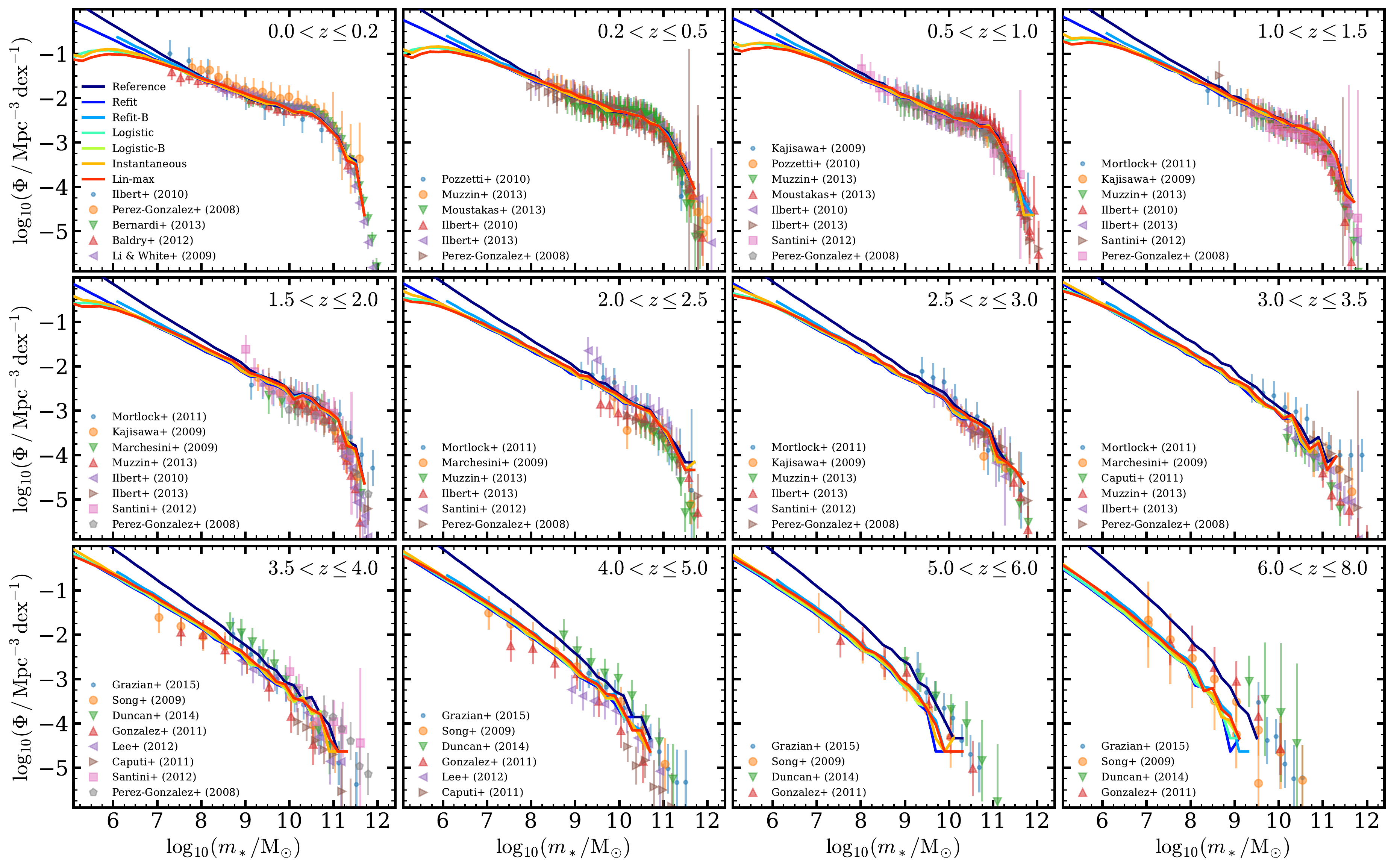}
	\centering
    \caption[]{The model predicted galaxy stellar mass function compared to the input observables for each of our model variants}
	\label{fig:app_smf}
\end{figure*}

\begin{figure*}[]
	\includegraphics[width=0.98\textwidth,keepaspectratio]{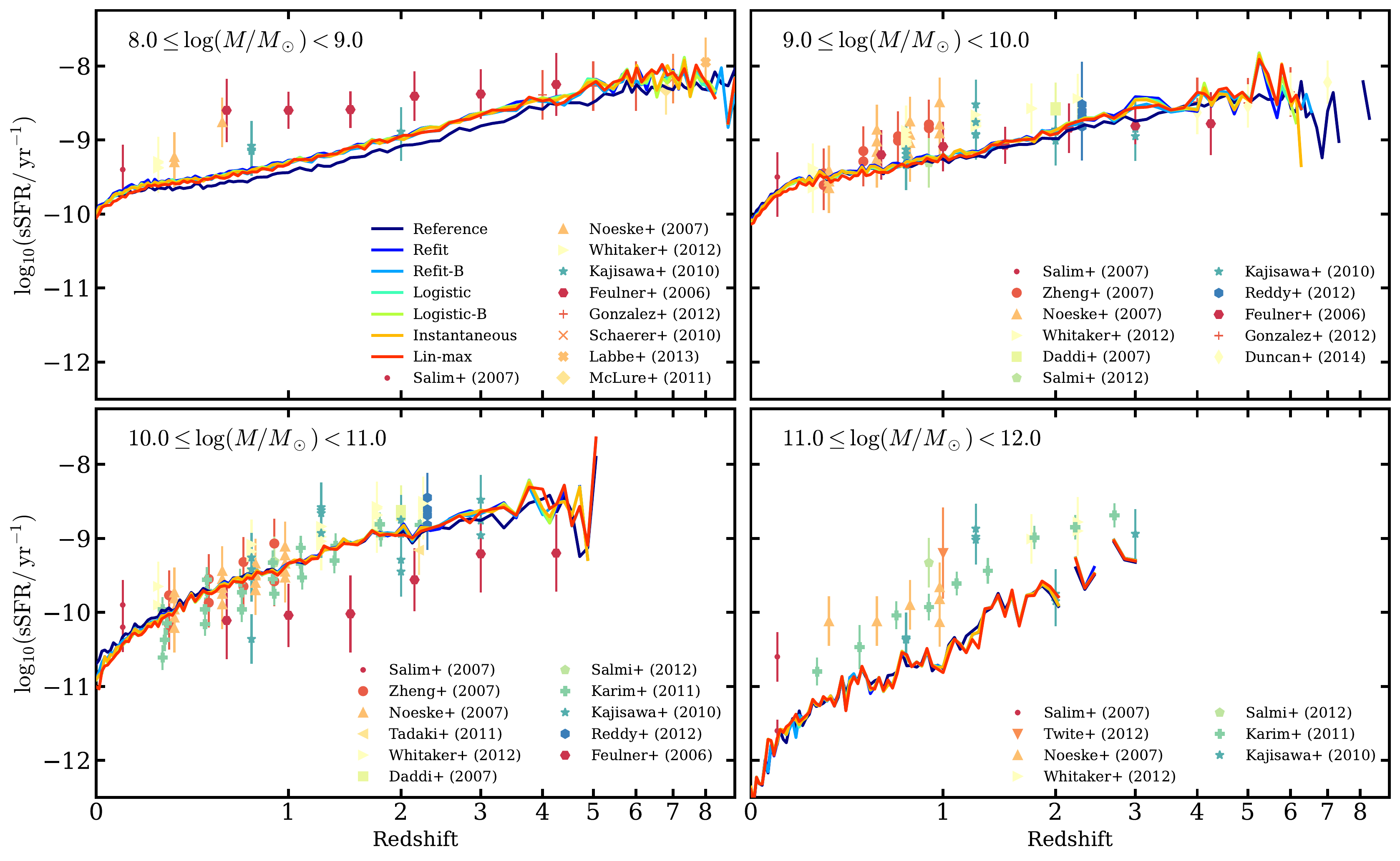}
	\centering
    \caption[]{The model predicted specific star formation rates compared to the input observables for each of our model variants}
	\label{fig:app_ssfr}
\end{figure*}

\begin{figure*}
	\includegraphics[width=0.98\textwidth,keepaspectratio]{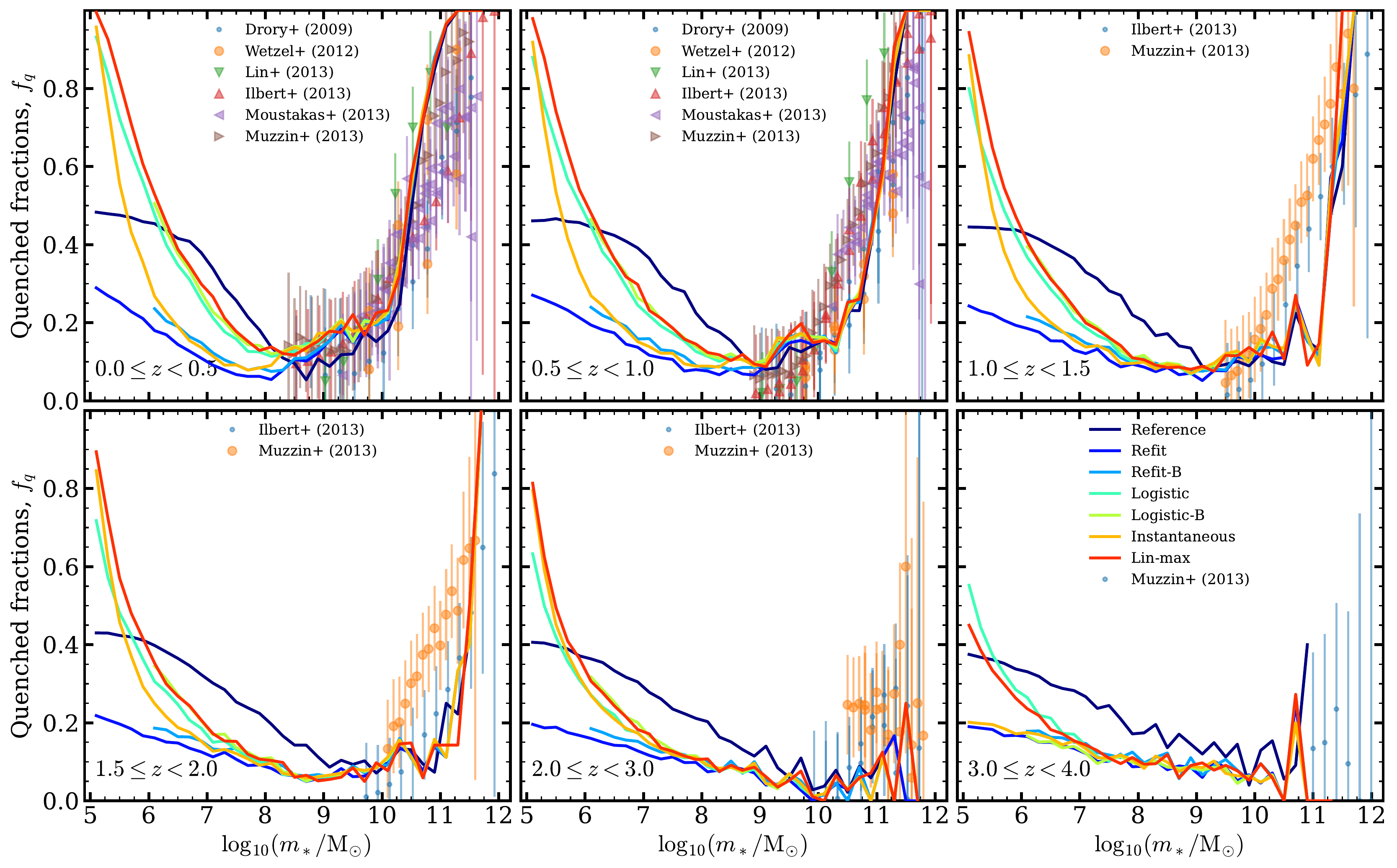}
	\centering
    \caption[]{The model predicted quenched fractions compared to the input observables for each of our model variants}
	\label{fig:app_fq}
\end{figure*}

\subsection{Model predictions for dwarfs}

Here, we show the complete results of our analysis for the model variants we did not highlight in the main text.

\begin{figure*}
	\includegraphics[width=0.9\textwidth, height=\textheight, keepaspectratio]{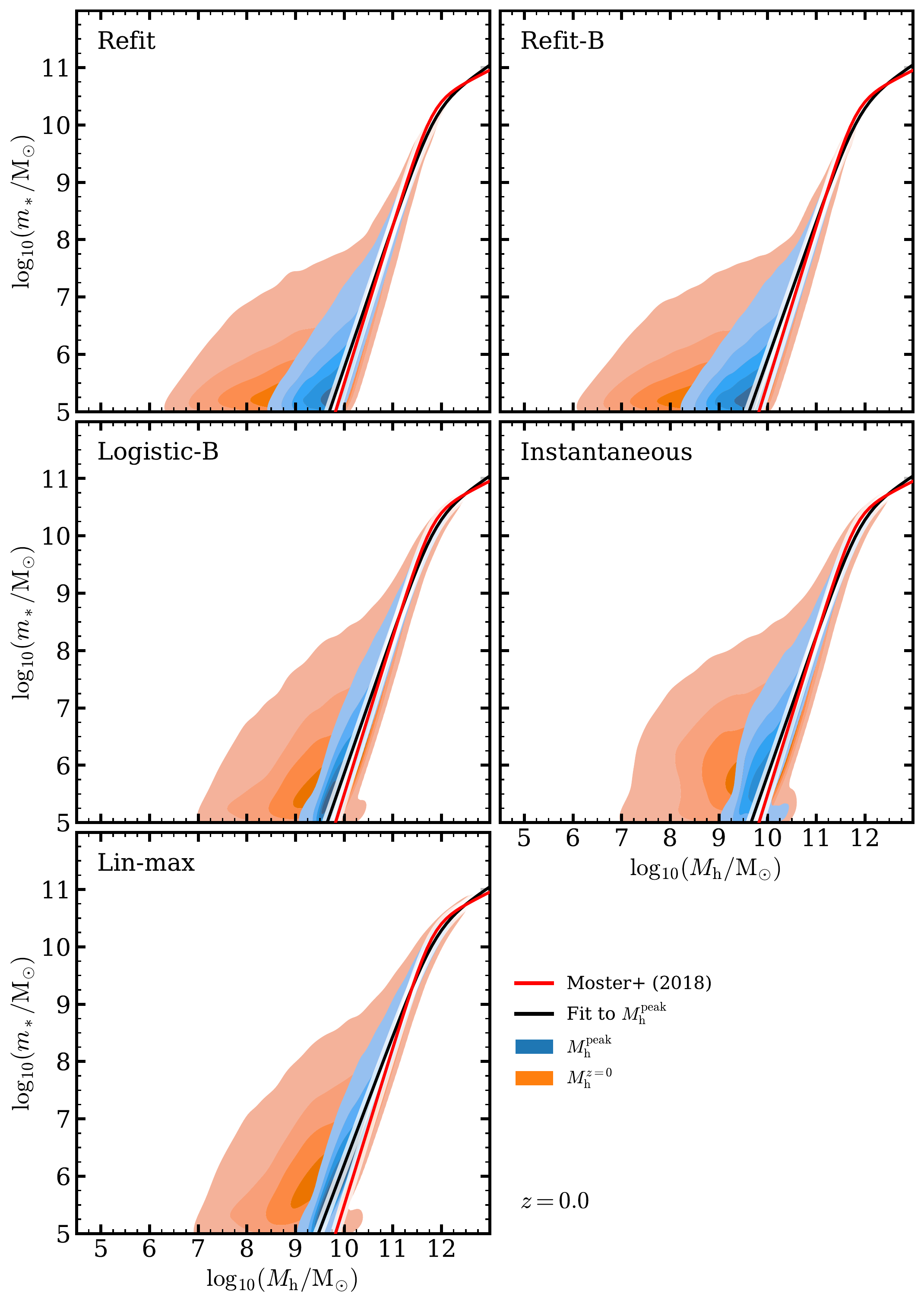}
	\centering
    \caption[]{The predicted SHMR for each model variant}
	\label{fig:app_shmr}
\end{figure*}

\begin{figure*}
	\includegraphics[width=0.98\textwidth,keepaspectratio]{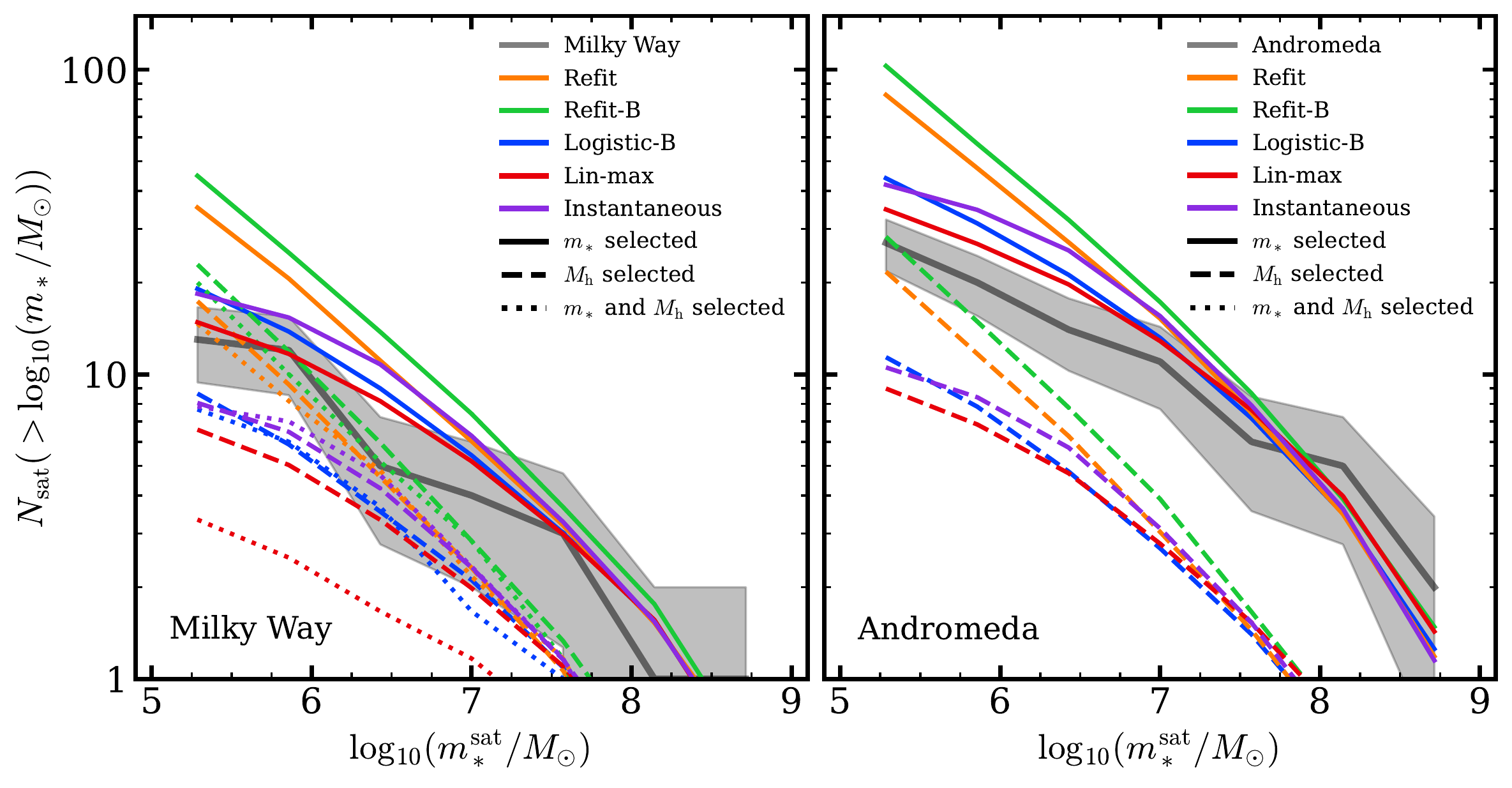}
	\centering
    \caption[]{The predicted cumulative dwarf satellite mass function for each model variant}
	\label{fig:app_cumulative_smf}
\end{figure*}

\begin{figure*}
	\includegraphics[width=0.98\textwidth,keepaspectratio]{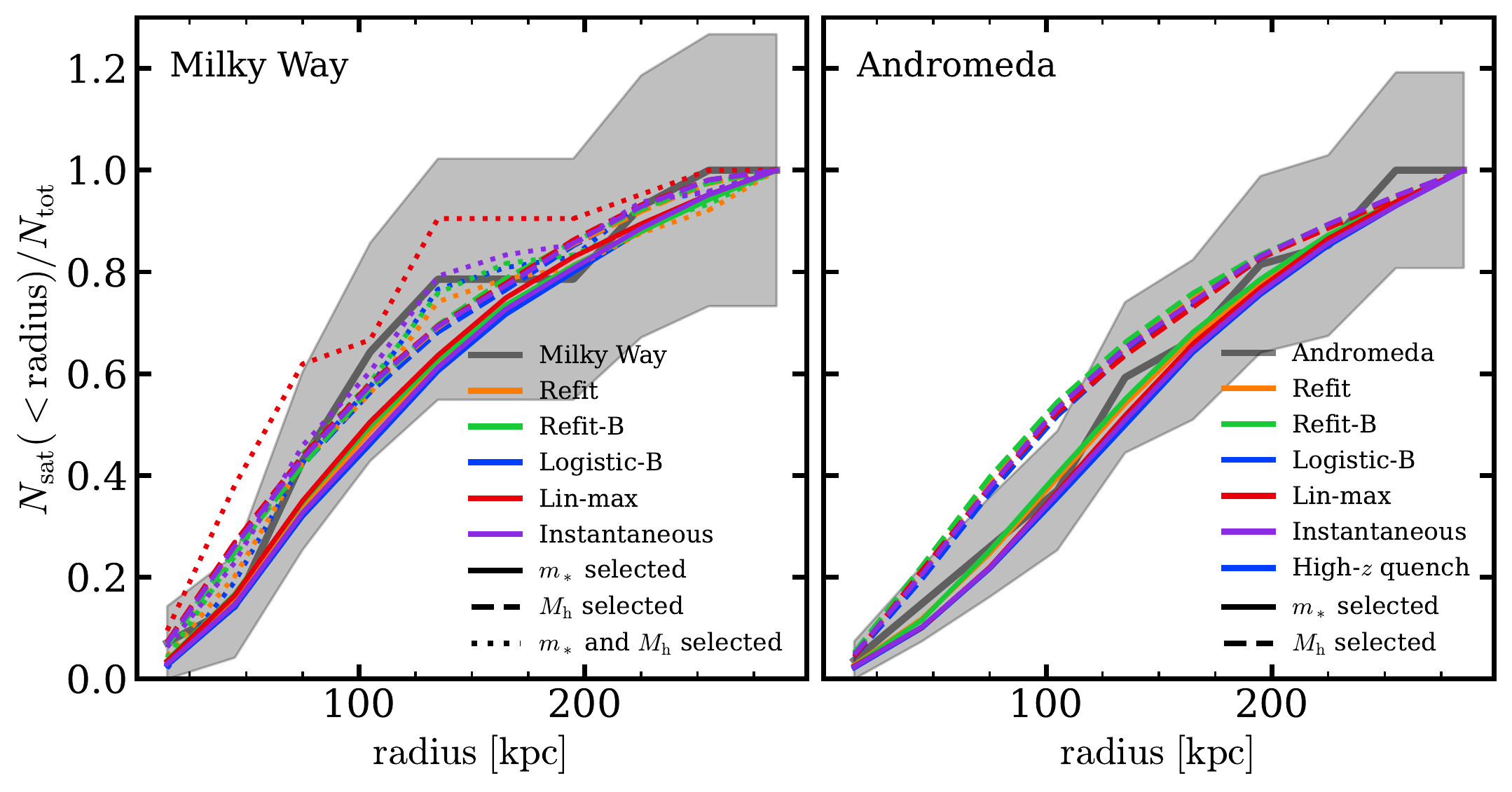}
	\centering
    \caption[]{The predicted cumulative dwarf satellite radial distribution for each model variant}
	\label{fig:app_cumulative_smf_radial}
\end{figure*}

\begin{figure*}
	\includegraphics[width=0.9\textwidth, keepaspectratio]{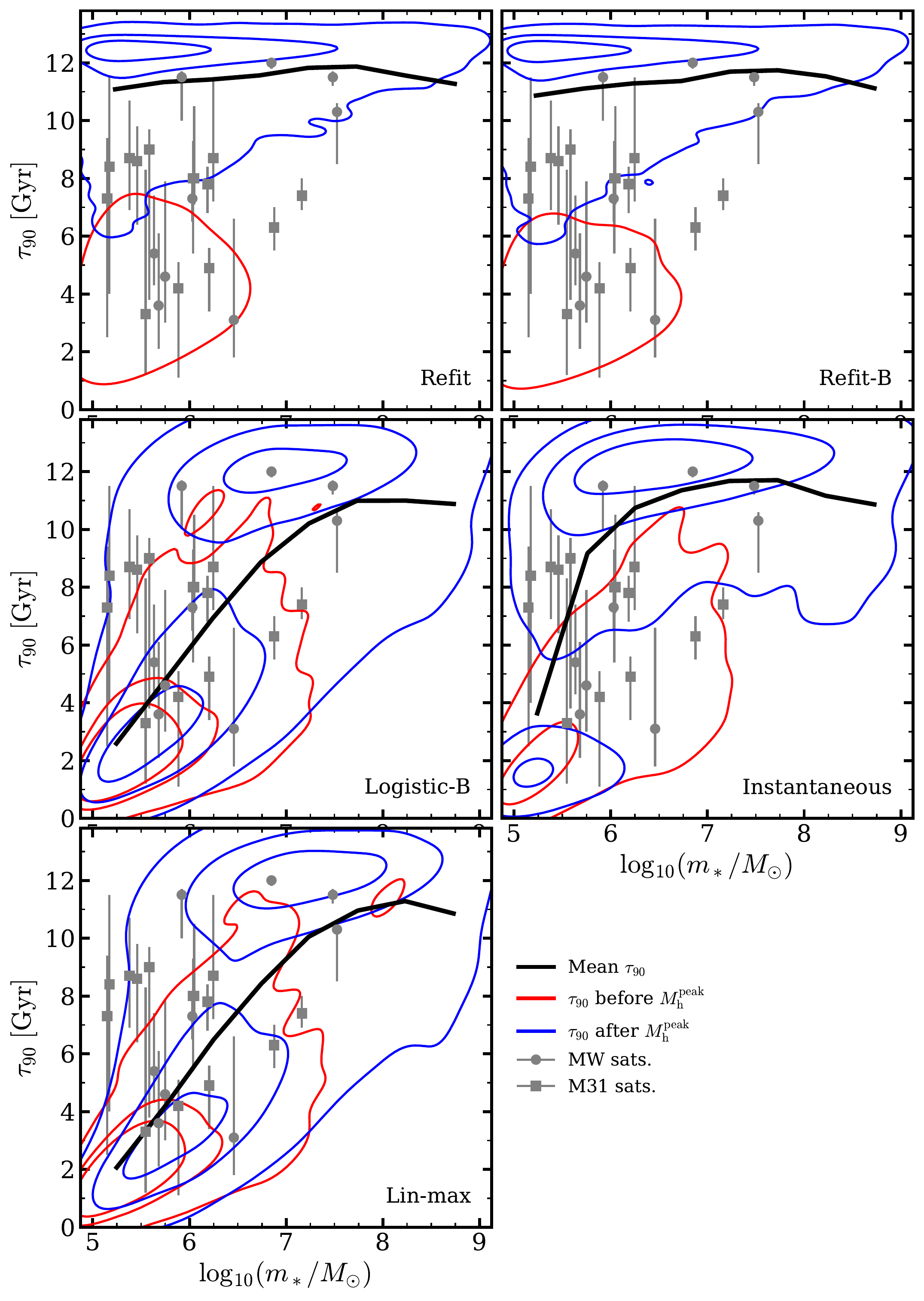}
	\centering
    \caption[]{The predicted $90$ per cent formation timescale for each model variant}
	\label{fig:app_t90}
\end{figure*}


\bsp	
\label{lastpage}
\end{document}